\documentclass[aps,prb,longbibliography,reprint,floatfix]{revtex4-2}

\usepackage[utf8]{inputenc}     
\usepackage[T1]{fontenc}        
\usepackage[english]{babel}     

\usepackage{amsmath}            

\usepackage{graphicx}           

\usepackage{multirow}           

\usepackage{xcolor}             
\definecolor{myblue}{HTML}{2EA0F0}  

\usepackage{hyperref}           
\hypersetup{
    colorlinks  = true,     
    linkcolor   = myblue,   
    citecolor   = myblue,   
    urlcolor    = myblue    
}

\usepackage{lineno}             

\input{mystyle}

\begin{document}

\title{Endorsing Titanium-Scandium Radionuclide Generator \texorpdfstring{\\}{ }
for PET and Positronium Imaging}

\author{\textsc{Pawe{\l} Moskal}\textsuperscript{\textit{a,b}}}
\author{\textsc{Aleksander Khreptak}\textsuperscript{\textit{a,b}}}
\email{aleksander.khreptak@uj.edu.pl}
\author{\textsc{Jaros{\l}aw Choi{\'n}ski}\textsuperscript{\textit{c}}}
\author{\textsc{Pete Jones}\textsuperscript{\textit{d}}}
\author{\textsc{Ihor Kadenko}\textsuperscript{\textit{e}}}
\author{\textsc{Agnieszka Majkowska-Pilip}\textsuperscript{\textit{f}}}
\author{\textsc{Rudrajyoti Palit}\textsuperscript{\textit{g}}}
\author{\textsc{Anna Stolarz}\textsuperscript{\textit{c}}}
\author{\textsc{Rafa{\l} Walczak}\textsuperscript{\textit{f}}}
\author{\textsc{Ewa St{\k{e}}pie{\'n}}\textsuperscript{\textit{a,b}}}

\affiliation{\textsuperscript{\textit{a}}M. Smoluchowski Institute of Physics, Jagiellonian University, Łojasiewicza 11, 30-348 Krakow, Poland}
\affiliation{\textsuperscript{\textit{b}}Center for Theranostics, Jagiellonian University, Kopernika 40, 31-501 Krakow, Poland}
\affiliation{\textsuperscript{\textit{c}}Heavy Ion Laboratory, University of Warsaw, Ludwika Pasteura 5A, 02-093 Warsaw, Poland}
\affiliation{\textsuperscript{\textit{d}}iThemba LABS, National Research Foundation, Old Faure Rd, Mfuleni, 7100 Cape Town, South Africa}
\affiliation{\textsuperscript{\textit{e}}Department of Nuclear and High Energy Physics, Taras Shevchenko National University of Kyiv, Volodymyrska Str. 64/13, 01601 Kyiv, Ukraine}
\affiliation{\textsuperscript{\textit{f}}Centre of Radiochemistry and Nuclear Chemistry, Institute of Nuclear Chemistry and Technology, Dorodna 16, 03-195 Warsaw, Poland}
\affiliation{\textsuperscript{\textit{g}}Department of Nuclear and Atomic Physics, Tata Institute of Fundamental Research, Dr Homi Bhabha Rd, Colaba, 400005 Mumbai, India}

\begin{abstract}
\noindent
The development of PET and positronium imaging techniques is strictly related to the availability of suitable radionuclides and robust radiochemistry platforms.
Among the emerging candidates, $^{44}$Sc has attracted significant interest due to its favourable physical properties, including a half-life of $\sim4$ hours, a pure ${\beta}^+$ emission profile, and the additional prompt $\gamma$-emission that enables advanced triple-photon detection schemes.
These characteristics make $^{44}$Sc particularly promising for high-resolution imaging and novel quantitative methodologies.
However, routine clinical and preclinical implementation requires a practical, sustainable, and cost-efficient production route.
In this context, we propose a titanium-scandium radionuclide generator as an optimal solution.
This study focuses on optimising the synthesis of the long-lived parent isotope, $^{44}$Ti (T$_{1/2}$~=~59.1 years), from which $^{44}$Sc can be selectively eluted in a chemically pure form when needed.
An analysis of various production pathways was conducted, including proton and deuteron reactions on scandium, as well as $\alpha$-particle and lithium-induced reactions on calcium, to determine the most efficient reaction parameters, target design, and expected yield.
Furthermore, we identify some existing cyclotron facilities suitable for implementing this technology.
Results indicate that efficient $^{44}$Ti production is achievable using proton beams in the 20--30~MeV range under extended irradiation conditions.
The proposed generator system would enable routine and decentralised $^{44}$Sc supply.
Its integration with the novel J-PET scanner may significantly reduce diagnostic costs and improve access to advanced PET imaging in regions with limited medical imaging infrastructure.

\vspace{2ex}
\noindent
topics: $^{44}$Ti/$^{44}$Sc generator, PET, positronium imaging, J-PET
\end{abstract}

\onecolumngrid
\begin{center}
    \textit{Proceedings of the 2nd Symposium on New Trends in Nuclear and Medical Physics}
\end{center}
\twocolumngrid

\maketitle


\section{Introduction}
\label{sec:intro}

Recently, the $^{44}$Sc radionuclide has been revisited as a promising candidate that may help democratise \textit{Positron Emission Tomography} (PET)~\cite{Moskal2024a} and enhance its specificity through \textit{positronium imaging}~\cite{Moskal2020, Moskal2021a, Moskal2024b}.

PET stands as a cornerstone of modern diagnostic imaging, providing unique, non-invasive insights into the molecular and cellular processes of the human body~\cite{Alavi2021}.
Since the first blurry images of brain tumours were obtained in the mid-twentieth century~\cite{Brownell1953}, PET has undergone a profound transformation, culminating in its clinical introduction in the 1970s~\cite{TerPogossian1975}.
This progress was driven by consistent advances in both radiation detection technologies~\cite{Jones2017} and radiopharmaceutical science~\cite{Clarke2018}, significantly strengthening its role in medical diagnostics.

A key advantage of PET, compared with established anatomical imaging techniques such as X-ray, Computed Tomography (CT), and Magnetic Resonance Imaging (MRI), is its ability to integrate functional data with anatomical context~\cite{Ryan2019}.
This capability to visualise metabolic processes \textit{in vivo} is essential for the early detection of disease, for monitoring metabolic changes over time, and for supporting precise, personalised diagnostics and therapeutic planning~\cite{Alavi2021, Bharathi2025}.

The operational principle of PET relies on the detection of $\gamma$ photons emitted from an administered radiopharmaceutical tracer, which accumulates in regions of high biochemical activity, typically corresponding to pathological lesions such as tumours~\cite{Berger2003}.
Consequently, PET provides essential data across numerous medical disciplines, notably in (i) oncology~\cite{Kirienko2024}, (ii) the diagnosis of neurological disorders (e.g., sclerosis, Parkinson’s, and Alzheimer’s disease)~\cite{Basu2009}, and (iii) cardiological assessments of heart function and blood flow~\cite{Rischpler2013}.

A recent and rapidly developing extension of this method is positronium imaging -- a technique that enables the direct visualisation of positronium characteristics within body tissues~\cite{Moskal2021a, Moskal2022, Bass2023, Moskal2024b, Moskal2025a, Parzych2023, Das2023, Takyu2022, Takyu2023}.
Positronium is a short-lived exotic atom formed by an electron and its antiparticle, the positron~\cite{Bass2023}.
Measurements of specific parameters, such as the \textit{ortho}-positronium mean lifetime, provide a novel biomarker of tissue microstructure~\cite{Moskal2023, Karimi2023}.
This biomarker offers information on features such as intermolecular pore size and the concentration of paramagnetic molecules, particularly oxygen~\cite{Shibuya2020, Moskal2021d}, thereby complementing conventional PET data on radiopharmaceutical uptake~\cite{Moskal2022, Das2025}.
The potential of this approach for enhancing diagnostic specificity at the molecular level has been demonstrated in the first successful \textit{ex vivo}~\cite{Moskal2021b} and \textit{in vivo}~\cite{Moskal2024b} studies using a dedicated multi-photon Jagiellonian PET (J-PET) scanner.
Positronium imaging is currently undergoing intensive development.
Beyond the J-PET scanner in Cracow, Poland \cite{Moskal2024b}, several other PET systems worldwide have the technical capability to measure the lifetime of positronium, including
the Prism-PET scanner in New York City, USA~\cite{Huang2025a},
the Biograph Vision Quadra in Bern, Switzerland~\cite{Prenosil2022, Steinberger2024, Mercolli2024},
the PennPET Explorer in Philadelphia, USA~\cite{Karp2020, Dai2023, Huang2024a, Huang2025b},
the NeuroEXPLORER (NX) brain PET scanner in New Haven, USA~\cite{Samanta2023}, and
the brain-dedicated TOF-PET scanner VRAIN in Chiba, Japan~\cite{Takyu2024a, Takyu2024b}.
Recent progress in developing and optimising reconstruction algorithms for \textit{Positronium Lifetime Imaging} (PLI)~\cite{Moskal2025a, Qi2022, Huang2024a, Huang2024b, Huang2024c, Huang2025a, Huang2025b, H_Huang2025, Berens2024, Chen2023, Chen2024, Shopa2023} is advancing the accuracy and reliability of positronium-based imaging.

The success of positronium imaging crucially depends on the choice of radionuclide.
An ideal isotope must possess two key attributes: (i) a half-life suitable for clinical protocols and (ii) a high probability of emitting a `\textit{prompt gamma}' quantum alongside positron decay, which serves as the essential time marker for positronium formation~\cite{Moskal2020, Das2025}.
While the $^{68}$Ga isotope (obtained from the widely available $^{68}$Ge/$^{68}$Ga generator~\cite{Nelson2022}) has been used in pioneering human positronium imaging trials~\cite{Moskal2024b}, its decay properties are suboptimal.
Specifically, only $\sim$1.34\% of $^{68}$Ga decays are accompanied by the required prompt $\gamma$ emission~\cite{Moskal2024b}, which complicates accurate measurements of positronium lifetime~\cite{Das2025}.
Consequently, $^{44}$Sc has attracted considerable attention among alternative radionuclides.
This isotope features a clinically suitable half-life T$_{1/2}$ of about 4~hours~\cite{NuDat3} and, crucially, emits a high-energy $\gamma$ quantum (1157~keV) in nearly 100\% of decays~\cite{NuDat3} almost immediately (2.6~ps) after positron emission~\cite{Sitarz2020}.
This decay profile makes $^{44}$Sc exceptionally suitable for advanced PET applications and positronium imaging.

The $^{44}$Sc isotope (T$_{1/2}$ = 3.97~h, or the recently reported 4.04~h~\cite{GarciaTorano2016, Duran2022}) presents a promising alternative to the widely used $^{68}$Ga (T$_{1/2}$ = 67.84 min~\cite{NuDat3}) for tumour PET and positronium imaging.
The nearly four times longer half-life of $^{44}$Sc enables the visualisation of tracer distribution over several hours post-injection.
This is particularly valuable for ligands with prolonged pharmacokinetics~\cite{Trencsenyi2023} and facilitates delayed imaging protocols and dynamic dosimetry~\cite{Eppard2017, Benabdallah2023}.
In addition, $^{44}$Sc emits positrons with a lower mean energy ($\sim$0.63~MeV) compared with $^{68}$Ga ($\sim$0.83~MeV)~\cite{Rosar2020}, resulting in a shorter positron range in tissue and thereby enhancing the spatial resolution of PET images by reducing blurring effects.
Moreover, its higher positron branching fraction ($\sim$94\% vs. $\sim$88\% for $^{68}$Ga~\cite{NuDat3}) increases the fraction of decays yielding a detectable positron.
An additional advantage lies in the chemical similarity of Sc(III) to the therapeutic radionuclide Lu(III)~\cite{MajkowskaPilip2011, Vaughn2021}.
This enables the use of identical targeting ligands to create matched molecular pairs: a diagnostic agent with $^{44}$Sc and a therapeutic counterpart based on either $^{177}$Lu~\cite{Umbricht2017} (a $\beta^-$ emitter, T$_{1/2}$~=~6.64 days~\cite{NuDat3}) or $^{47}$Sc~\cite{Domnanich2017} (a $\beta^-$ emitter, T$_{1/2}$~=~3.35 days~\cite{NuDat3}).
Together, these properties make scandium a unique theranostic platform.
The $^{44}$Sc/$^{47}$Sc pair, sharing identical coordination chemistry, exhibits matched biodistribution profiles \textit{in vivo}, enabling precise dosimetry for targeted radionuclide therapy with $^{47}$Sc via PET imaging with $^{44}$Sc.

The initial applications of $^{44}$Sc in nuclear medicine already demonstrate its effectiveness for precise tumour diagnostics.
One of the key strengths of $^{44}$Sc is its ability to form stable complexes with a wide range of chelators (e.g., DOTA, DTPA, EDTA, and NOTA)~\cite{Pruszynski2012, Kerdjoudj2016}, a versatility essential for labelling diverse biomolecules, including peptides and proteins. This adaptability is proving highly valuable across various diagnostic applications.
For example, $^{44}$Sc-labelled DOTATOC (a DOTA-conjugated peptide) has been successfully validated \textit{in vitro} and \textit{in vivo}, enabling high-quality visualisation of neuroendocrine tumours with excellent radiochemical purity and stability~\cite{Pruszynski2012, Singh2017}.
Similarly, $^{44}$Sc-PSMA-617 has demonstrated effectiveness in targeting prostate-specific membrane antigen in prostate cancer, offering high-contrast PET images essential for early and precise disease diagnosis~\cite{Umbricht2017, Eppard2017, Khawar2018}.
Furthermore, the successful evaluation of $^{44}$Sc-labelled Affibody molecules for imaging HER2-expressing tumours highlights the broad potential of this isotope in personalised medicine~\cite{Honarvar2017}, confirming that $^{44}$Sc-labelled ligands are a highly promising platform for the point-of-care diagnosis of various tumour types.

A recent milestone in the development of positronium imaging was the first experimental demonstration of PLI using $^{44}$Sc with the modular J-PET scanner~\cite{Das2025}. Conducted on a NEMA Image Quality phantom, this study validated the feasibility of using $^{44}$Sc as an optimal radionuclide for PLI.
The reconstructed images confirmed that $^{44}$Sc enables reliable identification of positronium-related events and accurate lifetime measurements, in close agreement with reference values in water.
Importantly, the plastic scintillator-based J-PET system~\cite{Moskal2020, Moskal2014, Moskal2021c, Sharma2023, Das2024, Ardebili2024}, operating in triggerless multi-photon detection mode~\cite{Korcyl2014, Korcyl2018, Moskal2021b, Moskal2025b}, demonstrated the capacity to separate positronium-specific signals from conventional PET events~\cite{Das2025}, thus establishing a solid foundation for future clinical translation of $^{44}$Sc-based positronium imaging.

Despite its clear advantages, the widespread adoption of $^{44}$Sc is complicated by challenges related to its availability.
Conventional short-lived positron-emitting nuclides for PET, such as $^{18}$F and $^{11}$C, are typically produced directly in cyclotrons~\cite{Pichler2018}, while others such as $^{68}$Ga are obtained from generators (e.g., the $^{68}$Ge/$^{68}$Ga system~\cite{Nelson2022}).
For $^{44}$Sc, both production routes are potentially feasible~\cite{Krajewski2013, Benabdallah2023, Schmidt2023, Larenkov2021, Gajecki2023}.

Direct cyclotron production of $^{44}$Sc has been successfully demonstrated in several studies~\cite{vanderMeulen2015,vanderMeulen2020, Krajewski2013, Dellepiane2024}.
The most efficient method identified is the $^{44}$Ca(p,n)$^{44}$Sc reaction using proton beams of 11--18 MeV on highly enriched $^{44}$Ca targets and yielding high activities with radionuclidic purity exceeding 99\%~\cite{vanderMeulen2015,vanderMeulen2020,Krajewski2013}.
As a practical alternative, the $^{47}$Ti(p,$\alpha$)$^{44}$Sc reaction has been successfully implemented on a medical cyclotron, offering simplified chemical processing of the final product~\cite{Dellepiane2024}.
While the direct cyclotron route enables centralised production at the tens-of-GBq level for distribution to clinical centres~\cite{vanderMeulen2015}, it faces practical limitations. These include the requirement for cyclotron access and enriched isotopic targets, along with the co-production of radionuclidic impurities such as $^{43}$Sc and $^{46}$Sc~\cite{Krajewski2013}.
Importantly, these impurities can be effectively minimised through optimised reaction parameters, particularly careful selection of proton energy and irradiation duration.

An alternative approach for supplying hospitals with $^{44}$Sc is the development of a titanium-scandium ($^{44}$Ti/$^{44}$Sc) generator system~\cite{Benabdallah2023}.
Similar to how the $^{68}$Ge/$^{68}$Ga generator made $^{68}$Ga widely available without cyclotrons~\cite{Nelson2022}, a generator based on the long-lived $^{44}$Ti could provide a continuous source of $^{44}$Sc for PET diagnostics and positronium imaging.

The parent isotope, $^{44}$Ti, has a long half-life of 59.1~years~\cite{NuDat3}.
It decays by electron capture to $^{44}$Sc, which in turn undergoes $\beta^+$ decay, emitting annihilation photons (511~keV) and a prompt 1157~keV gamma quantum.
A single generator of this type could potentially supply a hospital with $^{44}$Sc for many years of daily use~\cite{Moskal2024a}.
This solution is particularly valuable for medical facilities without access to an on-site cyclotron or in settings requiring regular production of radiopharmaceuticals with short-lived isotopes.

The principal challenge, however, lies in producing $^{44}$Ti itself in sufficient quantities.
The $^{45}$Sc(p,2n)$^{44}$Ti reaction, achieved by irradiating metallic scandium with high-energy protons, represents the most established production route~\cite{Ejnisman1996, Hassan2018, Schmidt2023}.
Nevertheless, other pathways, such as deuteron irradiation of scandium, and $\alpha$-particle~\cite{Hassan2018} or $^7$Li-induced reactions on calcium-based targets represent important alternative methods (see Sect.~\ref{sec:reactions}).
Due to the very long half-life of $^{44}$Ti, saturation of these reactions is practically unattainable within a feasible irradiation period~\cite{Schmidt2023}. The activity accumulates approximately linearly, necessitating many weeks of continuous operation of a high-power accelerator, coupled with an efficient radiochemical separation process, to obtain even hundreds of MBq of $^{44}$Ti.

Another significant challenge involves developing a $^{44}$Ti/$^{44}$Sc generator system that exhibits high performance in the separation of the parent-daughter pair.
This requires efficient retention of the parent Ti(IV) on the generator matrix and reproducible, quantitative elution of the daughter Sc(III), while minimising $^{44}$Ti breakthrough~\cite{Benabdallah2023}.

Addressing these challenges would enable the implementation of titanium-scandium generators as a dependable source of $^{44}$Sc for both PET and positronium imaging, thereby providing a powerful novel tool for personalised nuclear medicine.

\section{Nuclear reactions for \texorpdfstring{$^{44}$Ti}{44Ti} production}
\label{sec:reactions}

The successful implementation of the $^{44}$Ti/$^{44}$Sc generator depends on producing the long-lived parent isotope, $^{44}$Ti, in quantities sufficient for long-term generator operation via charged-particle-induced nuclear reactions.
The most promising channels involve proton, deuteron, and $\alpha$-particle reactions on targets of scandium-45 ($^{45}$Sc) or various calcium isotopes.
Proton-induced reactions are well-established experimentally and provide relatively high cross-sections~\cite{Ejnisman1996, Daraban2009, Hassan2018}.
While deuteron channels offer an alternative route, their reported yield is extremely low~\cite{Hassan2018}, making them significantly less efficient.
In contrast, $\alpha$-particle reactions open up a different spectrum of nuclear processes, but typically require higher-power accelerators and are accompanied by more complex isotopic impurity co-production~\cite{Hassan2018}.
In addition to these classical routes, reactions induced by $^7$Li ions on $^{40}$Ca have been explored as a potential source of $^{44}$Ti.

This section summarises the fundamental data and findings for these three reaction groups, forming a basis for optimising $^{44}$Ti synthesis conditions for future generator applications.

\subsection{Proton activation of \texorpdfstring{$^{45}$Sc}{45Sc}}
\label{sec:reactions-proton}

Experimental implementation of the $^{45}$Sc(p,2n)$^{44}$Ti reaction is facilitated by two key factors: (i) the availability of high-intensity proton beams in the required energy range on commercial cyclotrons, and (ii) the fact that natural scandium is monoisotopic ($^{45}$Sc)~\cite{NISTAtomicWeights}, which eliminates the need for expensive isotopic enrichment.
Analysis of the excitation function by McGee \textit{et al.}~\cite{McGee1970} suggests that the reaction proceeds predominantly through the compound nucleus mechanism.
In this model, a proton with energy above the threshold ($E_{\mathrm{thr}} \approx12.65$~MeV~\cite{Daraban2009}) is captured by $^{45}$Sc, forming an excited $^{46}$Ti$^*$ nucleus, which subsequently de-excites via two-neutron evaporation, yielding $^{44}$Ti.

\begin{figure}[!htbp]
    \centering
    \includegraphics[width=\linewidth]{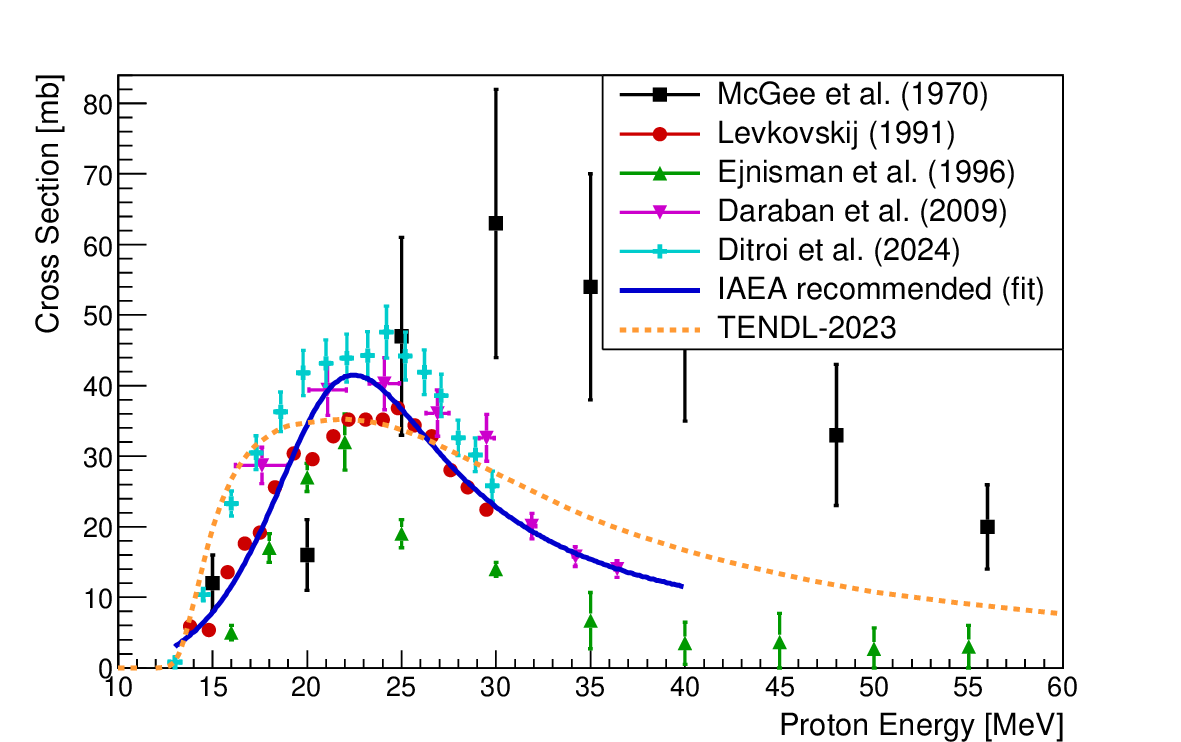}
    \caption{Experimental excitation function data for the $^{45}$Sc(p,2n)$^{44}$Ti reaction, including studies by McGee \textit{et al.}~\cite{McGee1970} (black squares), Levkovskij~\cite{Levkovskij1991} (red circles, normalised by a factor of 0.8), Ejnisman \textit{et al.}~\cite{Ejnisman1996} (green upward triangles), Daraban \textit{et al.}~\cite{Daraban2009} (magenta downward triangles), and Ditr{\'o}i \textit{et al.}~\cite{Ditroi2024} (cyan crosses).
    The IAEA database recommendation~\cite{IAEA-Medical} (solid blue line), based on a Pad{\'e} fit to evaluated data, and the TENDL-2023 evaluation~\cite{TENDL2023} (dotted orange line) are also presented.}
    \label{fig:xs_45Sc_p2n}
\end{figure}

The excitation function for the $^{45}$Sc(p,2n)$^{44}$Ti reaction, as visualised in Fig.~\ref{fig:xs_45Sc_p2n}, shows substantial variations across different studies. This is particularly evident in the reported maximum cross-section ($\sigma_{\mathrm{max}}$) values (see Table~\ref{tab:sigma-max}).
For example, measurements by McGee \textit{et al.}~\cite{McGee1970} report $\sigma_{\mathrm{max}} = 63 \pm 19$~mb at proton energy $E_{\mathrm{p}} \sim 30$~MeV, whereas later studies obtained lower maxima: $32 \pm 4$~mb at 22~MeV by Ejnisman \textit{et al.}~\cite{Ejnisman1996} and $47.6 \pm 3.7$~mb at 24.2~MeV by Ditr{\'o}i \textit{et al.}~\cite{Ditroi2024}.
Consequently, even for the peak region, the cross-section values vary by almost a factor of two.
This discrepancy can be attributed to variations in experimental methodologies, including differences in target thickness, beam current, detector calibration methods, and corrections for $\gamma$-ray detection efficiency.

In addition to experimental studies, theoretical evaluations are available in the TALYS Evaluated Nuclear Data Library (TENDL)~\cite{Koning2019}.
The TENDL-2023 data~\cite{TENDL2023} show certain discrepancies with experimental measurements, particularly in the cross-section magnitude and the energy location of the maximum (see Fig.~\ref{fig:xs_45Sc_p2n}).
Attempts to improve the theoretical agreement with experimental data through parameter variation are presented in the Appendix~\ref{app:model_parameters}.

For consistency and comparative analysis, the reference Charged-particle Cross Section Database for Medical Radioisotope Production (IAEA)~\cite{IAEA-Medical} provides recommended evaluated cross-section $\sigma(E_{\mathrm{p}})$ values based on a review of existing experiments and model calculations (see Table~\ref{tab:sigma-recom}).

\begin{table}[!htbp]
    \caption{Reported maximum cross-sections for the $^{45}$Sc$(p,2n)^{44}$Ti reaction from selected studies. 
    The table summarises the proton energy $E_{\mathrm{p}}$ at which the peak occurs and the corresponding maximum cross-section value $\sigma_{\mathrm{max}}$, 
    demonstrating the spread of reported results.}
    \label{tab:sigma-max}
    \centering
    \begin{tabular}{l|c|c}
    \hline\hline
    Reference & $E_{\mathrm{p}}$ [MeV] at peak & $\sigma_{\mathrm{max}}$ [mb] \\
    \hline
    McGee \textit{et al.}~\cite{McGee1970}   & 30 & 63 $\pm$ 19 \\
    Levkovskij~\cite{Levkovskij1991}$^{\dagger}$    & 24.8 & 36.8\,(corrected) \\
    Ejnisman \textit{et al.}~\cite{Ejnisman1996} & 22 & 32 $\pm$ 4 \\
    Daraban \textit{et al.}~\cite{Daraban2009}   & 24.1 & 40.3 $\pm$ 3.7 \\
    Ditr{\'o}i \textit{et al.}~\cite{Ditroi2024}   & 24.2 & 47.6 $\pm$ 3.7 \\
    IAEA Database~\cite{IAEA-Medical}   & 22.5 & 41.5 $\pm$ 4.4 \\
    TENDL-2023~\cite{TENDL2023} & 22 & 35.2 \\
    \end{tabular}

    \begin{flushleft}
    $^{\dagger}$~Values of Levkovskij were normalised by a factor of 0.8,  
    since the original monitor reaction data were found to be too high~\cite{Daraban2009}.
    \end{flushleft}
\end{table}

\begin{table*}[!htbp]
    \caption{Recommended cross sections ($\sigma(E_{\mathrm{p}})$) for the $^{45}$Sc(p,2n)$^{44}$Ti reaction from the IAEA database (Charged-Particle Cross-Section Database for Medical Radioisotope Production, identifier: \texttt{scp44ti0}).~\cite{IAEA-Medical}. The table presents proton energies $E_{\mathrm{p}}$ and corresponding recommended $\sigma$ values with their uncertainties.}
    \label{tab:sigma-recom}
    \centering
    \begin{tabular}{c|c|c|c|c|c|c|c}
    \hline\hline
    $E_{\mathrm{p}}$ & $\sigma$ & $E_{\mathrm{p}}$ & $\sigma$ & $E_{\mathrm{p}}$ & $\sigma$ & $E_{\mathrm{p}}$ & $\sigma$ \\
    {}[MeV] & [mb] & [MeV] & [mb] & [MeV] & [mb] & [MeV] & [mb] \\
    \hline
    13.0 & 3.0 $\pm$ 1.6 & 19.5 & 31.6 $\pm$ 3.6 & 26.0 & 33.5 $\pm$ 4.3 & 32.5 & 18.5 $\pm$ 2.9 \\
    13.5 & 4.0 $\pm$ 2.1 & 20.0 & 34.4 $\pm$ 3.6 & 26.5 & 32.0 $\pm$ 4.1 & 33.0 & 17.8 $\pm$ 2.8 \\
    14.0 & 5.1 $\pm$ 1.7 & 20.5 & 37.0 $\pm$ 3.9 & 27.0 & 30.5 $\pm$ 3.9 & 33.5 & 17.1 $\pm$ 2.7 \\
    14.5 & 6.4 $\pm$ 2.1 & 21.0 & 39.0 $\pm$ 4.0 & 27.5 & 28.6 $\pm$ 3.7 & 34.0 & 16.5 $\pm$ 3.2 \\
    15.0 & 7.9 $\pm$ 2.0 & 21.5 & 40.4 $\pm$ 4.2 & 28.0 & 27.6 $\pm$ 3.5 & 34.5 & 15.9 $\pm$ 3.1 \\
    15.5 & 9.6 $\pm$ 2.5 & 22.0 & 41.0 $\pm$ 4.3 & 28.5 & 26.8 $\pm$ 3.5 & 35.0 & 15.4 $\pm$ 3.1 \\
    16.0 & 11.6 $\pm$ 2.4 & 22.5 & 41.5 $\pm$ 4.4 & 29.0 & 25.1 $\pm$ 3.2 & 35.5 & 14.9 $\pm$ 2.9 \\
    16.5 & 13.9 $\pm$ 2.9 & 23.0 & 41.2 $\pm$ 4.4 & 29.5 & 23.6 $\pm$ 3.1 & 36.0 & 14.4 $\pm$ 3.2 \\
    17.0 & 16.4 $\pm$ 2.8 & 23.5 & 40.5 $\pm$ 4.6 & 30.0 & 22.0 $\pm$ 3.1 & 36.5 & 13.9 $\pm$ 3.4 \\
    17.5 & 19.2 $\pm$ 3.3 & 24.0 & 39.4 $\pm$ 4.7 & 30.5 & 21.8 $\pm$ 2.9 & 37.0 & 13.5 $\pm$ 3.2 \\
    18.0 & 22.1 $\pm$ 3.1 & 24.5 & 38.1 $\pm$ 4.6 & 31.0 & 20.9 $\pm$ 2.8 & 37.5 & 13.1 $\pm$ 3.2 \\
    18.5 & 25.3 $\pm$ 3.5 & 25.0 & 36.7 $\pm$ 4.6 & 31.5 & 20.0 $\pm$ 2.7 & 38.0 & 12.8 $\pm$ 3.7 \\
    19.0 & 28.5 $\pm$ 3.3 & 25.5 & 35.1 $\pm$ 4.4 & 32.0 & 19.2 $\pm$ 3.0 & 38.5 & 12.4 $\pm$ 3.6 \\
    \end{tabular}
\end{table*}

The excitation function exhibits characteristic energy dependence: the first reaction events appear just above threshold (13-16~MeV) with small cross-sections, followed by a rapid increase to a maximum in the 22-25~MeV range, and decreases beyond 35~MeV due to the opening of competing reaction channels.

\begin{figure}[!htbp]
    \centering
    \includegraphics[width=\linewidth]{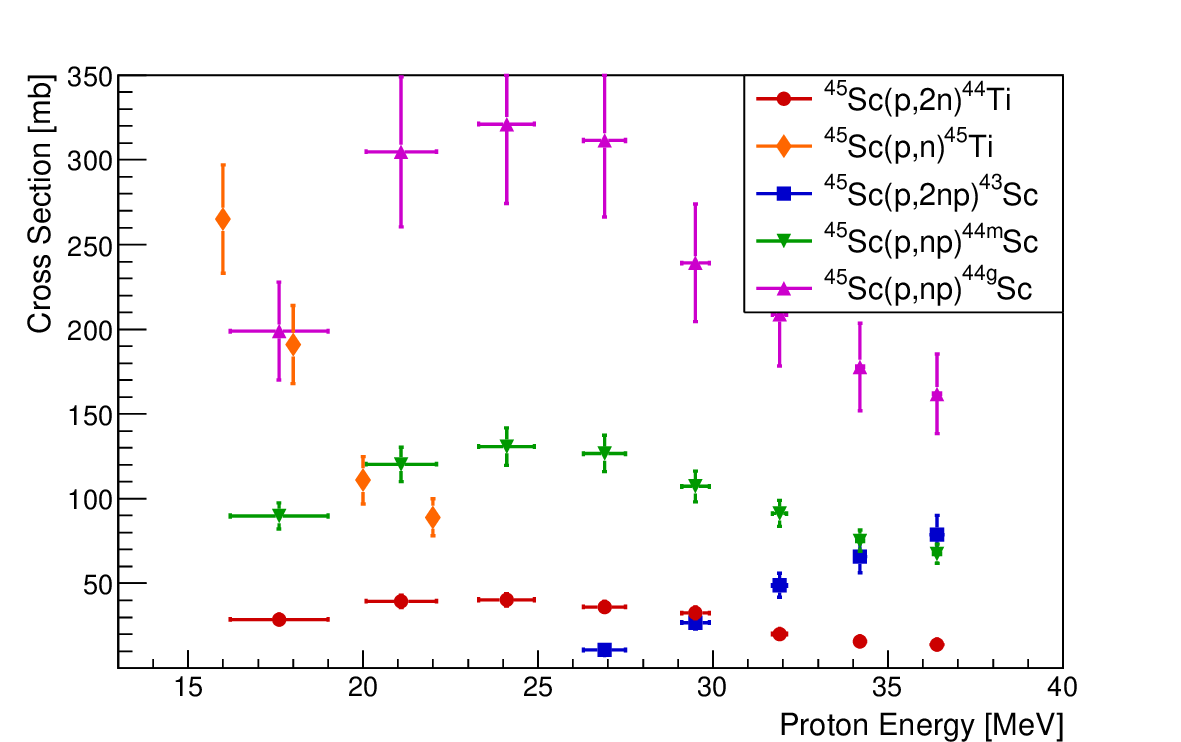}
    \caption{Experimentally measured cross-sections $\sigma$ for proton-induced nuclear reactions on $^{45}$Sc as a function of incident proton energy $E_{\mathrm{p}}$.
    The data show the excitation functions for the main $^{45}$Sc(p,2n)$^{44}$Ti production channel (red circles) and its primary competing reactions: $^{45}$Sc(p,n)$^{45}$Ti (orange diamonds), $^{45}$Sc(p,2np)$^{43}$Sc (blue squares), $^{45}$Sc(p,np)$^{44g}$Sc (green upward triangles), and $^{45}$Sc(p,np)$^{44m}$Sc (magenta downward triangles).
    Data from Daraban \textit{et al.}~\cite{Daraban2009} (scandium isotopes and $^{44}\text{Ti}$) and Ejnisman \textit{et al.}~\cite{Ejnisman1996} ($^{45}\text{Ti}$).}
    \label{fig:xs_45Sc_px}
\end{figure}

The production of $^{44}$Ti via the $^{45}$Sc(p,2n)$^{44}$Ti reaction proceeds alongside several competing nuclear processes, as illustrated in Fig.~\ref{fig:xs_45Sc_px}.
The most significant competitor is the $^{45}$Sc(p,np)$^{44}$Sc reaction, which has a similar threshold ($\sim$11.57~MeV) and exhibits cross-sections several times larger than the (p,2n) channel~\cite{Ejnisman1996, Daraban2009}.
This route produces the short-lived $^{44g}$Sc ($T_{1/2}$ = 4.04~h~\cite{NuDat3}) and its metastable isomer $^{44m}$Sc ($T_{1/2}$ = 58.62~h~\cite{NuDat3}), both $\beta^+$ emitters that create radionuclidic impurities and complicate direct use of the irradiated product.
However, these isotopes decay to negligible levels within a few days after irradiation and thus have no impact on the long-term use of a $^{44}$Ti/$^{44}$Sc generator.
The $^{45}$Sc(p,n)$^{45}$Ti, characterised by a low threshold ($\sim2.9$~MeV)~\cite{Kuhn2015, Hassan2018}, represents another important competing channel.
Although $^{45}$Ti ($T_{1/2}\approx$ 3.1~h~\cite{NuDat3}) decays via $\beta^+$ emission to stable $^{45}$Sc and does not persist in the final $^{44}$Ti product, its formation during irradiation consumes proton flux, thereby reducing the overall $^{44}$Ti production yield.
At higher energies (>25~MeV), multi-particle emission channels become significant~\cite{Daraban2009}: $^{45}$Sc(p,2np)$^{43}$Sc ($T_{1/2} \approx$ 3.9~h, $\beta^+$~\cite{NuDat3}) and $^{45}$Sc(p,3p)$^{43}$K ($T_{1/2} \approx$ 22.3~h, $\beta^-$~\cite{NuDat3}).
These reactions generate impurities that undergo radioactive decay during and after target processing, thereby influencing the overall radionuclidic purity.

In practice, the radionuclidic impurity problem is addressed using a combined strategy, which consists of selecting the appropriate irradiation energy range and subsequent radiochemical purification.
For high-current irradiations of solid metallic $^{45}$Sc targets, the optimal energy window is between 16 and~30 MeV.
Within this range, the (p,2n) reaction achieves its maximum while the (p,n) channel decreases sufficiently and multi-nucleon emission processes have not yet become dominant.

\subsection{Deuteron activation of \texorpdfstring{$^{45}$Sc}{45Sc}}
\label{sec:reactions-deuteron}

\begin{figure}[!htbp]
    \centering
    \includegraphics[width=\linewidth]{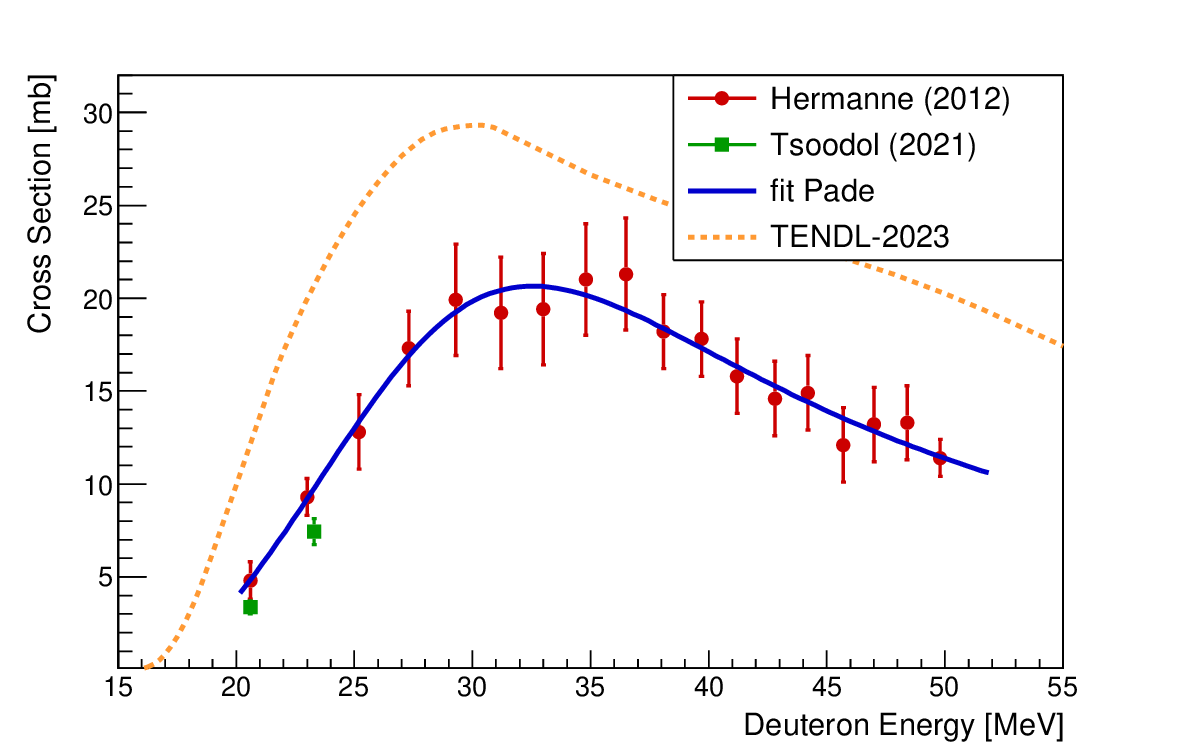}
    \caption{Excitation function for the $^{45}$Sc(d,3n)$^{44}$Ti reaction. Experimental cross-section data from Hermanne \textit{et al.}~\cite{Hermanne2012} (red circles) and Tsoodol \textit{et al.}~\cite{Tsoodol2021} (green squares) are shown alongside a Pad{\'e} fit (solid blue line) and the TENDL-2023 evaluation (dotted orange line)~\cite{TENDL2023}.
    The plot illustrates the relatively low cross-sections for this channel, peaking in the 30--37~MeV range.}
    \label{fig:xs_45Sc_d}
\end{figure}

The deuteron-induced $^{45}$Sc(d,3n)$^{44}$Ti reaction provides an alternative route for $^{44}$Ti production.
This process involves deuteron absorption followed by the subsequent emission of three neutrons from the compound nucleus.
The reaction has a threshold of $\sim$15.3 MeV~\cite{Hassan2018}, and its excitation function is shown in Fig.~\ref{fig:xs_45Sc_d}. 
Experimental data indicate that measurable $^{44}$Ti yields appear above 20~MeV~\cite{Hermanne2012, Tsoodol2021}, with a broad maximum in the 30--37~MeV range.
The reported cross-sections are relatively low, reaching approximately 19--21~mb, and the fitted excitation function peaking at around 32--33~MeV.
The TENDL-2023~\cite{TENDL2023} predicts different behaviour: the calculated cross-sections are significantly higher than the experimental values, and the maximum of the excitation function is shifted towards lower deuteron energies.
An analysis of parameter sensitivity in theoretical cross-section calculations is provided in the Appendix~\ref{app:model_parameters}.
In comparison with the proton-induced $^{45}$Sc(p,2n)$^{44}$Ti reaction, the deuteron channel exhibits both a higher threshold and significantly lower cross-sections, making it less attractive for large-scale $^{44}$Ti production.

\begin{figure}[!htbp]
    \centering
    \includegraphics[width=\linewidth]{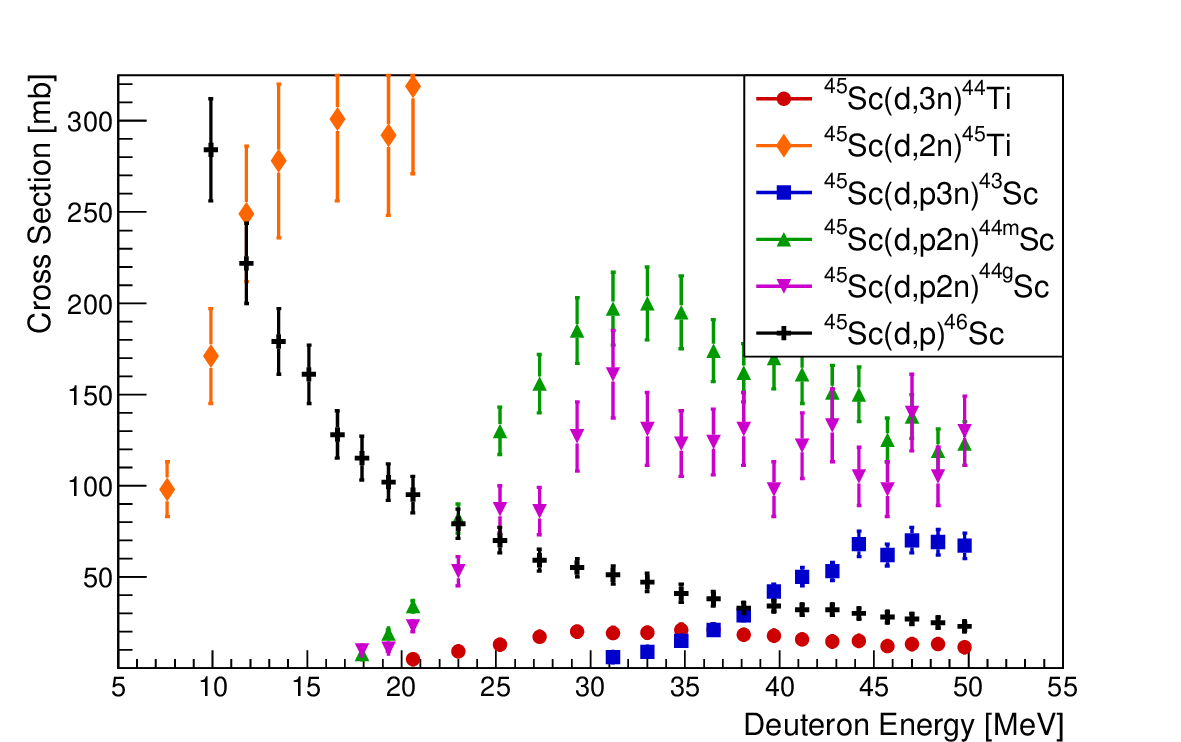}
    \caption{Excitation functions for deuteron-induced reactions on $^{45}$Sc, showing the main channel $^{45}$Sc(d,3n)$^{44}$Ti (red circles) alongside competing reactions: $^{45}$Sc(d,2n)$^{45}$Ti (orange diamonds), $^{45}$Sc(d,p3n)$^{43}$Sc (blue squares), $^{45}$Sc(d,p2n)$^{44m}$Sc (green upward triangles), $^{45}$Sc(d,p2n)$^{44g}$Sc (magenta downward triangles), and $^{45}$Sc(d,p)$^{46}$Sc (black crosses).
    Data from Hermanne \textit{et al.}~\cite{Hermanne2012}.}
    \label{fig:xs_45Sc_d_all}
\end{figure}

During deuteron irradiation of $^{45}$Sc, the production of $^{44}$Ti via the $^{45}$Sc(d,3n) channel competes with several significant nuclear reactions~\cite{Hermanne2012, Tsoodol2021} that affect both yield and purity.
Some of these are shown in Fig.~\ref{fig:xs_45Sc_d_all}.
The most prominent competitor is the $^{45}$Sc(d,p2n)$^{44}$Sc reaction, which leads to the formation of $^{44g}$Sc and its metastable isomer $^{44m}$Sc. This channel exhibits cross-sections an order of magnitude higher ($\sim$100--200~mb) than the main $^{44}$Ti production route~\cite{Hermanne2012} (see Fig.~\ref{fig:xs_45Sc_d_all}).
Other notable competing channels include $^{45}$Sc(d,2n)$^{45}$Ti, $^{45}$Sc(d,p3n)$^{43}$Sc, and $^{45}$Sc(d,p)$^{46}$Sc, the latter being a long-lived ($T_{1/2}$ = 83.8~days~\cite{NuDat3}) impurity produced efficiently at relatively low deuteron energies~\cite{Skobelev2011}.

Whereas most radioactive impurities ($^{45}$Ti, $^{44}$Sc, $^{43}$Sc) decay to negligible levels within days after irradiation due to their relatively short half-lives, $^{46}$Sc presents a particular challenge for radionuclidic purity, as its long half-life and efficient formation introduce significant background $\gamma$-activity.
In addition, the stable $^{46}$Ti~\cite{Tsoodol2021}, produced both directly via the high-cross-section $^{45}$Sc(d,n) reaction and indirectly via $\beta^-$ decay of $^{46}$Sc~\cite{NuDat3}, represents a significant non-radioactive impurity.
This channel converts a substantial part of the $^{45}$Sc target without contributing to the desired $^{44}$Ti product.

Overall, these competing reactions reduce the effective $^{44}$Ti yield and complicate both spectroscopic analysis and subsequent chemical separation processes due to the complex mixture of radioactive and stable isotopes.
Consequently, the optimal deuteron energy range lies between 30 and 35~MeV.
Within this range, the $^{45}$Sc(d,3n)$^{44}$Ti reaction approaches its maximum cross-section, while the production of long-lived impurities, particularly $^{46}$Sc from the (d,p) channel, is significantly reduced.
However, the persistent dominance of competing (d,2n) and (d,n) channels requires thorough post-irradiation chemical purification to achieve acceptable radionuclidic purity.

\subsection{Irradiation of calcium isotopes with \texorpdfstring{$\alpha$}{alpha} particles}
\label{sec:reactions-alpha}

Alpha-particle irradiation of calcium isotopes provides an alternative route for $^{44}$Ti production.
In this approach, fusion of the incident helium nucleus (Z~=~2) with the calcium target nucleus (Z~=~20) directly yields the titanium compound nucleus (Z~=~22).

The $^{40}$Ca($\alpha$,$\gamma$)$^{44}$Ti reaction represents a direct production route via radiative capture.
This exothermic process (Q~=~+5.13~MeV~\cite{Mohr2015}) involves the formation of an excited $^{44}$Ti$^*$ compound nucleus, which subsequently de-excites by $\gamma$ emission.
With no energy threshold, the reaction occurs even at low $\alpha$-particle energies, where it proceeds through narrow resonances~\cite{Mohr2019}.
However, cross-sections for this channel are extremely small, of the order of microbarns~\cite{Mohr2015}.
Above approximately 6~MeV, particle-emission channels, such as ($\alpha$,p), open and quickly dominate, suppressing the ($\alpha$,$\gamma$) yield~\cite{Mohr2019}.
As a result, absolute $^{44}$Ti yields remain very low, which causes significant challenges for practical production of $^{44}$Ti for radionuclide generator applications.

In addition to radiative capture, $^{44}$Ti can be produced via multi-neutron evaporation following $\alpha$-irradiation of calcium isotopes: $^A$Ca($\alpha$,xn)$^{44}$Ti, where $A$~=~42, 43, 44 and x~=~2, 3, 4, respectively.

\begin{figure}[!htbp]
    \centering
    \includegraphics[width=\linewidth]{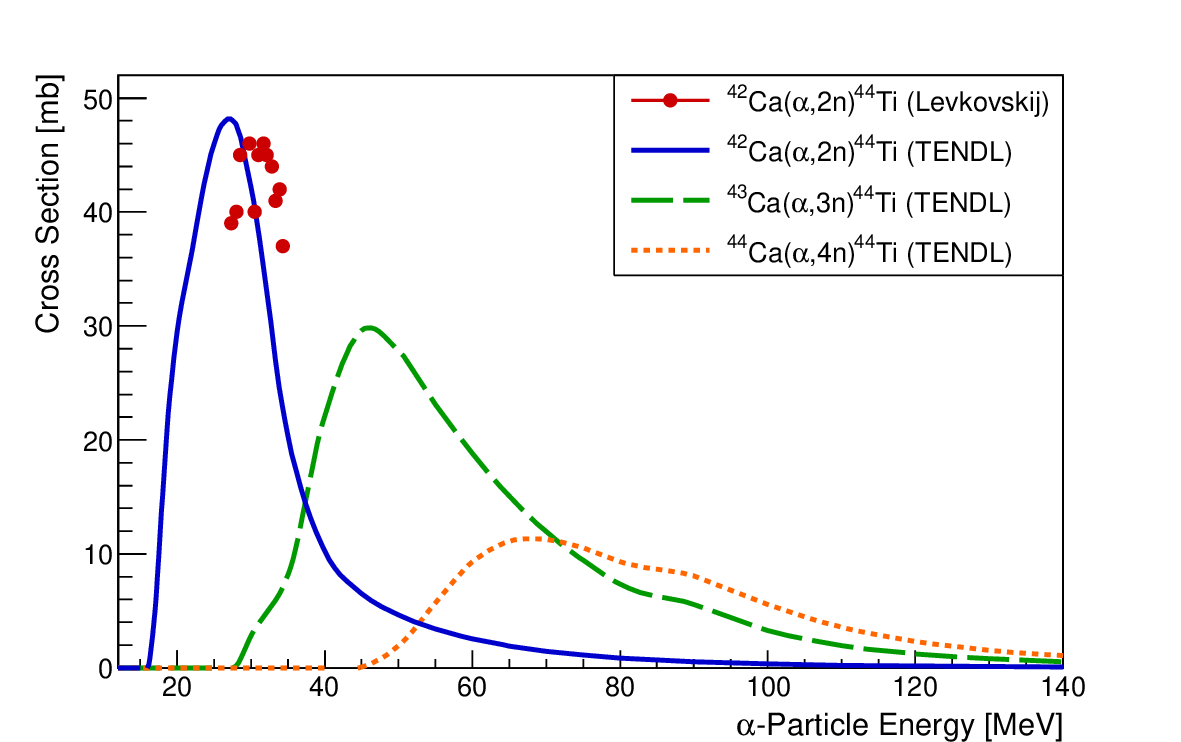}
    \caption{Theoretical excitation functions of the $^{42}$Ca($\alpha$,2n)$^{44}$Ti (solid blue line), $^{43}$Ca($\alpha$,3n)$^{44}$Ti (dashed green line), and $^{44}$Ca($\alpha$,4n)$^{44}$Ti (dotted orange line) reactions as predicted by the TENDL-2023 library~\cite{TENDL2023}.
    The experimental data point for $^{42}$Ca reaction (red circles) is from Levkovskij~\cite{Levkovskij1991}.}
    \label{fig:xs_Ca_alpha}
\end{figure}

For $^{42}$Ca($\alpha$,2n)$^{44}$Ti, cross-sections of $\sim$40--46~mb have been reported in the energy range $E_\alpha \sim$27--34~MeV~\cite{Levkovskij1991} (Fig.~\ref{fig:xs_Ca_alpha}), indicating that measurable yields are achievable only at relatively high $\alpha$-particle energies.
However, this channel competes strongly with $^{42}$Ca($\alpha$,n)$^{45}$Ti~\cite{Cheng1979} and $^{42}$Ca($\alpha$,p)$^{45}$Sc~\cite{Buckby1983}, both of which are open at lower energies and reach $\sigma$ on the order of $\mathcal{O}(10^2)$~mb.

Even higher beam energies are required for $\alpha$-induced reactions on $^{43,44}$Ca isotopes, with thresholds at 24.9~MeV and 36.9~MeV, respectively~\cite{Hassan2018}.
Since no experimental cross-sections are available for these reactions, their excitation functions were evaluated using the TENDL-2023 data library~\cite{TENDL2023}.
As shown in Fig.~\ref{fig:xs_Ca_alpha}, the calculated cross-section for the $^{43}$Ca($\alpha$,3n)$^{44}$Ti reaction peaks at $\sim$30~mb around 45~MeV, whereas the $^{44}$Ca($\alpha$,4n)$^{44}$Ti channel reaches a lower maximum of $\sim$11~mb near 65~MeV.
Additional theoretical excitation functions under varying model parameters are presented in Appendix~\ref{app:model_parameters}.
Importantly, the $^{45}$Ti-producing reactions (e.g. $^{43}$Ca($\alpha$,2n)$^{45}$Ti) have lower thresholds and their excitation functions overlap extensively with the $^{44}$Ti-producing channels over a broad interval ($\sim$25--70 MeV)~\cite{Hassan2018}.

Finally, natural isotopic composition constrain target choice: $^{42}$Ca, $^{43}$Ca and $^{44}$Ca occur at about 0.647\%, 0.135\% and 2.086\% of natural calcium, respectively~\cite{NISTAtomicWeights}, so substantial production would require isotopic enrichment.
Furthermore, accelerators capable of delivering $\alpha$-particle beams at suitable energies (e.g., >30 MeV) are relatively rare compared to those used for proton or deuteron bombardment.

\subsection{\texorpdfstring{$^{7}$Li}{7Li}-induced reactions on \texorpdfstring{$^{40}$Ca}{40Ca}}
\label{sec:reactions-lithium}

The irradiation of $^{40}$Ca with $^{7}$Li ions enables the production of $^{44}$Ti through several reaction channels.
The dominant mechanism is the $^{40}$Ca($^{7}$Li,t)$^{44}$Ti reaction via cluster transfer, where the weakly-bound $^{7}$Li nucleus (separating into $\alpha$ + t clusters) dissociates upon interaction with the target~\cite{Cutler1980}.
This results in the $\alpha$-particle fusing with $^{40}$Ca to form $^{44}$Ti, while the triton is emitted.
The low binding energy of $^{7}$Li in the t~+~$\alpha$ channel ($\sim$2.47~MeV~\cite{Chen2020}) favours this cluster mechanism.
Alternative pathways, such as $^{40}$Ca($^{7}$Li,p2n)$^{44}$Ti and $^{40}$Ca($^{7}$Li,dn)$^{44}$Ti, proceed via complete fusion forming a $^{47}$V compound nucleus, followed by evaporation of multiple particles -- either individual nucleons or nucleon-cluster combinations (e.g., d~+~n).
These processes require higher excitation energies to overcome particle separation thresholds.
Numerical simulations of the $^{40}$Ca($^{7}$Li,x)$^{44}$Ti reaction cross-sections were performed, with the resulting excitation functions presented and discussed in Appendix~\ref{app:model_parameters}.
Significant competing reactions that reduce radionuclidic purity include: neutron evaporation leading to $^{46}$V, proton emission yielding $^{46}$Ti, and emission of light particles such as deuterons or $^3$He, potentially producing isotopes like $^{45}$Ti and $^{44}$Sc.

The primary challenge of this approach is the low selectivity of $^{44}$Ti, as competing reaction channels dominate the reaction flux and complicate subsequent chemical separation.

\vspace{2ex}
In summary, proton irradiation of natural scandium represents the most viable method for $^{44}$Ti production towards a $^{44}$Ti/$^{44}$Sc generator system, with deuteron-, alpha-, and lithium-induced reactions being less favourable for practical implementation due to lower yields and greater technical constraints.

\section{Targets}
\label{sec:targets}

Depending on the chosen or available projectile, target materials for $^{44}$Ti production include natural scandium ($^{45}$Sc) for proton or deuteron bombardment, and calcium-based targets for irradiation with $\alpha$ particles or $^7$Li ions.
Considering the relatively low cross-sections for the reactions leading to $^{44}$Ti (Figs.~\ref{fig:xs_45Sc_p2n}, \ref{fig:xs_45Sc_d}, and \ref{fig:xs_Ca_alpha}), the target thickness must be carefully optimised.
It should be tailored to the specific energy window where the cross-section for the desired reaction is significant.
The use of thicker targets should be avoided to minimise the production of competing radionuclidic impurities (as discussed in Sect.~\ref{sec:reactions}), which are produced at lower or higher energies.

The target thicknesses were calculated for the projectile energy window in which the cross-section for the desired reaction remains significant.
This was achieved by integrating the inverse mass stopping power, $S(E)$, from the selected incident energy, $E_{\mathrm{in}}$, down to an \textit{effective} lower energy limit, $E_{\mathrm{out}}$, below which the reaction cross-section becomes negligible~\cite{IAEA_TRS465}:

\begin{equation}
    \label{eq:thickness}
    x = \int_{E_{\mathrm{out}}}^{E_{\mathrm{in}}}\frac{1}{S(E)}\,dE,
\end{equation}

\noindent
where $x$ denotes the resulting areal density of the target material, expressed in mg/cm$^{2}$.
The stopping power data were obtained from SRIM (Stopping and Range of Ions in Matter) software package~\cite{Ziegler2010, SRIM} for the relevant particle-target pairs: protons and deuterons in scandium, and $\alpha$ particles and $^7$Li ions in CaCO$_3$.
The calculated areal densities for various energy windows are summarised in Table~\ref{tab:thickness}.

For metallic targets, this quantity can be converted to a linear thickness $d$ using the relation $d = x / \rho$, where $\rho$ is the material's density.
For example, a scandium target ($\rho_{\mathrm{Sc}}$ = 2.989~g/cm$^{3}$~\cite{CRC2019}) with an areal density corresponding to the 16.5$\to$12~MeV proton energy window yields a physical thickness of approximately 640~$\mu$m, while the 30$\to$12~MeV interval requires about 3.5~mm.
In the case of deuteron beams, the required thickness increases from about 310~$\mu$m at 20~MeV to nearly 5~mm at 50~MeV.

\begin{table}[!htbp]
    \caption{Calculated target thickness required to decelerate the projectile from $E_{\mathrm{in}}$ down to the chosen $E_{\mathrm{out}}$, evaluated using Eq.~(\ref{eq:thickness}) and stopping powers from SRIM~\cite{Ziegler2010, SRIM}.}
    \label{tab:thickness}
    \centering
    \begin{tabular}{c|c|c|c|c}
    \hline\hline
    Projectile & Target & $E_{\mathrm{out}}$ [MeV] & $E_{\mathrm{in}}$ [MeV] & $x$ [mg/cm$^{2}$] \\
    \hline
    \multirow{1}{*}{$p$} & \multirow{1}{*}{$^{\mathrm{nat}}$Sc} & \multirow{1}{*}{12} 
    & 16.5 & 191 \\
    &  &  & 20 & 371 \\
    &  &  & 30 & 1025 \\
    \hline
    \multirow{1}{*}{$d$} & \multirow{1}{*}{$^{\mathrm{nat}}$Sc} & \multirow{1}{*}{17} 
    & 20 & 93 \\
    &  &  & 30 & 478 \\
    &  &  & 50 & 1581 \\
    \hline
    \multirow{1}{*}{$\alpha$} & \multirow{1}{*}{CaCO$_3$} & \multirow{1}{*}{27.5} 
    & 35 & 41 \\
    &  &  & 50 & 146 \\
    &  &  & 70 & 331 \\
    \hline
    \multirow{1}{*}{$^{7}$Li} & \multirow{1}{*}{CaCO$_3$} & \multirow{1}{*}{13}
    & 25 & 13 \\
    & & & 35 & 28 \\
    & & & 50 & 58 \\
    \end{tabular}
\end{table}

Scandium is a relatively soft metal, which facilitates its mechanical processing. Targets can be prepared by reshaping bulk material through rolling or by pressing metal pieces into a pellet format using a hydraulic press.
As scandium slowly covers with an oxide layer in the ambient atmosphere, it is recommended that these procedures be carried out in an inert atmosphere.

For production routes involving calcium, the chemical form of the isotopically enriched material is a crucial consideration. 
The enriched isotopes are typically supplied as calcium carbonate (CaCO$_3$).
While conversion to the metallic form (Ca) is possible~\cite{Stinson1976}, this process presents practical challenges.
The conversion efficiency is typically around 80--85\%, leading to a non-negligible loss of the valuable enriched material.
Furthermore, unless a direct vacuum transfer line exists between the conversion apparatus and the irradiation station, the highly reactive metallic calcium will rapidly oxidise in air, reforming calcium carbonate and negating the conversion.

These issues can be avoided by preparing the target directly from the calcium carbonate.
However, in this case, the problem of the heat dissipation deposited by the beam increases dramatically, since carbonates are thermal insulators~\cite{Stolarz2020}.
A potential solution is forming the target with the addition of a thermally conductive, chemically inert material such as carbon~\cite{Stolarz2018}.
This can be achieved by creating a homogeneous mixture of CaCO$_3$ and carbon powder, whereas a sandwich-type format would be less practicable given the required geometric thickness constraints.

\section{Yield calculation}

The thick-target yield (TTY) formalism, which integrates both the nuclear reaction cross-section and the stopping power of the target material~\cite{Aikawa2015}, is used to estimate production yields.
The TTY, $Y$ (in nuclei per incident particle), is given by:

\begin{equation}
    \label{eq:tty}
    Y = \frac{N_A}{M}\int_{E_{\mathrm{out}}}^{E_{\mathrm{in}}}\frac{\sigma(E)}{S(E)}\,dE,
\end{equation}

\noindent
where $N_A$ is the Avogadro's constant and $M$ is the molar mass of the target material (e.g., $\sim$44.96~g/mol for $^{45}$Sc~\cite{NISTAtomicWeights}).

For a beam of particles with charge state $q$ and current $I$, the rate of incident protons is 
$\Phi = I/(qe)$, where $e$ is the elementary charge.
The production rate of $^{44}$Ti nuclei is then:

\begin{equation}
    \label{eq:rate}
    R = \Phi Y = \frac{I}{q e}\,Y.
\end{equation}

Taking into account radioactive decay during the irradiation, the activity at 
the end of bombardment (EOB) is:

\begin{equation}
    \label{eq:activity_eob}
    A_{\mathrm{EOB}}(I,t) = R \left(1 - e^{-\lambda t}\right) = \frac{I}{q e}\,Y \left(1 - e^{-\lambda t}\right),
\end{equation}

\noindent
where $\lambda = \ln 2/T_{1/2}$ is the decay constant of $^{44}$Ti.

Given the relatively long half-life of $^{44}$Ti ($T_{1/2}$~=~59.1~years~\cite{NuDat3}), the product $\lambda t$ remains $\ll 1$ for practical irradiation times $t$ (hours to weeks), and thus radioactive decay during irradiation is negligible.
This allows the linear approximation
$ 1-e^{-\lambda t} \approx \lambda t$
to be applied to Eq.~\ref{eq:activity_eob}, simplifying the expression for the EOB activity to:

\begin{equation}
    \label{eq:activity_simple}
    A_{\mathrm{EOB}}(I,t) \approx \frac{I}{q e} Y \lambda t.
\end{equation}

\begin{figure}[!htbp]
    \centering
    \includegraphics[width=\linewidth]{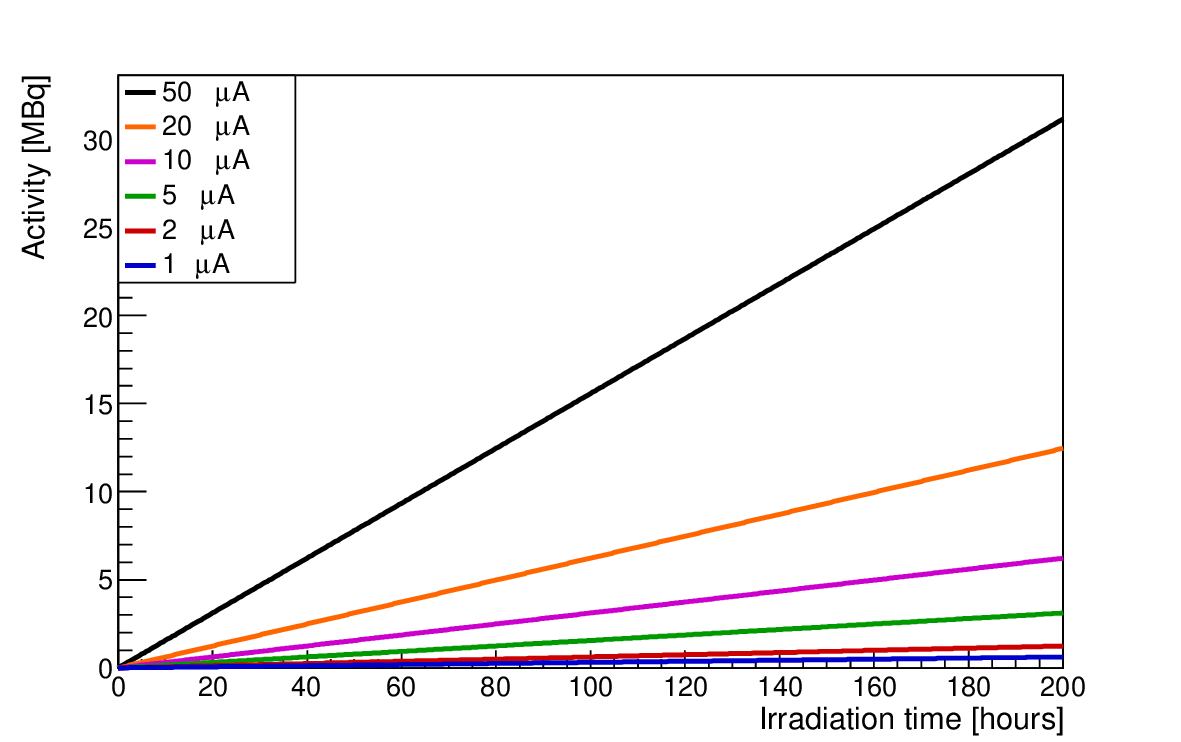}
    \caption{Accumulation of $^{44}$Ti activity at the end of bombardment (EOB) as a function of irradiation time for the $^{45}$Sc(p,2n)$^{44}$Ti reaction with an incident proton energy of 30~MeV.
    The linear growth is shown for different beam currents, illustrating the dependence on both time and beam intensity. The calculation assumes a thick scandium target, and the yield $Y$ was evaluated from Eq.~(\ref{eq:tty}) based on the cross-section data reported in~\cite{IAEA-Medical} and stopping power values calculated using SRIM~\cite{Ziegler2010, SRIM}.}
    \label{fig:activity_eob}
\end{figure}

Consequently, the activity accumulated at EOB depends approximately linearly on the irradiation time, as illustrated in Fig.~\ref{fig:activity_eob}.
Reaction saturation is therefore unattainable under typical laboratory conditions.

The integrated $^{44}$Ti yield scales linearly with the particle fluence. Thus, increasing the beam current provides a direct method for enhancing the production yield.
However, higher beam currents also significantly increase the thermal load on the target.
Effective cooling is essential to prevent target damage through melting or deformation.

A successful experimental implementation of the $^{45}$Sc(p,2n)$^{44}$Ti production route was demonstrated by Filosofov \textit{et al.}~\cite{Filosofov2010}.
The irradiation of a natural scandium target ($\sim$1.5~g) with 25~MeV protons yielded $\sim$185~MBq of $^{44}$Ti.
Subsequent radiochemical processing enabled creation of a $^{44}$Ti/$^{44}$Sc generator that showed efficient $^{44}$Sc elution with high parent-daughter separation.
This confirms the practical efficiency of the methodology from production to application.

\section{Post-irradiation processing and \texorpdfstring{$^{44}$Ti}{44Ti} separation}

Post-irradiation processing of targets is a crucial stage that determines the purity of the final $^{44}$Ti product and its suitability for subsequent use, particularly as the parent nuclide in $^{44}$Ti/$^{44}$Sc generators.

Comprehensive radiometric characterisation precedes any chemical separation of the irradiated target. High-resolution $\gamma$-ray spectrometry with high-purity germanium (HPGe) detectors serves as the principal analytical method, enabling precise identification of radionuclides through their characteristic $\gamma$ emissions and subsequent activity quantification.
The activity of $^{44}$Ti can be determined either by direct measurement of its characteristic low-energy gamma emissions (primarily 67.9~keV and 78.3~keV) or indirectly through the 1157~keV gamma line from its $^{44}$Sc daughter, measured after secular equilibrium is established~\cite{Filosofov2010, Radchenko2016, Radchenko2017, Lee2017}.
The $\gamma$ spectrum is simultaneously analysed for characteristic emissions of co-produced radionuclidic impurities, which exhibit distinct spectral signatures. 
This preliminary radiometric assessment provides important data for evaluating production yield and determining the optimal strategy for subsequent chemical purification processes.

The chemical processing of irradiated metallic scandium targets commonly uses dissolution in concentrated mineral acids, such as hydrochloric acid (HCl).
This step converts the bulk target material into soluble Sc(III) ions, while the produced $^{44}$Ti is present in solution as Ti(IV).
Literature reports varying conditions for this procedure.
For example, Filosofov \textit{et al.}~\cite{Filosofov2010} used 2 M HCl for complete target dissolution, while Radchenko \textit{et al.} applied higher concentrations (6--12.3~M HCl) to ensure complete dissolution and solution stability~\cite{Radchenko2016, Radchenko2017}.

The separation of $^{44}$Ti relies on the differential ion-exchange behaviour of Ti(IV) and Sc(III) in a chloride medium.
An initial separation is typically achieved by loading the solution onto an anion-exchange column in concentrated HCl, where titanium chlorocomplexes are retained and scandium is eluted.
This step reduces the scandium matrix and achieves a Ti/Sc separation factor of 10$^2$–10$^3$~\cite{Radchenko2017}.
A subsequent fine purification using cation-exchange chromatography in HCl removes residual scandium, resulting in a $^{44}$Ti recovery of 90--97\% and overall Ti/Sc separation exceeding 10$^6$~\cite{Filosofov2010, Radchenko2017}.

Solid-phase extraction chromatography offers an efficient alternative to ion-exchange methods~\cite{Radchenko2016, Benabdallah2023}.
Hydroxamate-based ZR resin, for instance, demonstrates exceptional sorption affinity for Ti(IV) across a wide range of HCl concentrations (0.1--10~M), with distribution coefficients ($K_{\mathrm{d}}$) exceeding 1000, while showing minimal retention of Sc(III) ($K_{\mathrm{d}} < 3$)~\cite{Radchenko2016}.
This high selectivity enables efficient separation even from large target matrices.
Titanium can subsequently be eluted using complexing agents such as HCl/H$_2$O$_2$ mixtures, oxalic acid, or citric acid.
In contrast, branched diglycolamide (BDGA) resin exhibits an inverse retention profile, strongly binding Sc(III) ($K_{\mathrm{d}} \sim$730 at 4~M HCl), while Ti(IV) shows negligible retention ($K_{\mathrm{d}} \leq 1$ below 6~M HCl)~\cite{Radchenko2016}.
This characteristic makes BDGA particularly suitable for the fine purification of $^{44}$Ti, enabling selective removal of residual scandium while allowing titanium to pass through the column.

Detailed procedures for extracting $^{44}$Ti from CaCO$_3$ targets irradiated with either $\alpha$ particles or $^7$Li ions are not well documented.
The production of $^{44}$Ti occurs in extremely low quantities within a massive calcium matrix, making its chemical separation challenging.
By analogy with established methods for $^{43,44}$Sc recovery from CaCO$_3$ targets~\cite{Krajewski2013, Becker2023}, a potential approach may involve initial dissolution of the irradiated target in a concentrated mineral acid (e.g., HCl) under heating (typically 80$^{\circ}$C), followed by chromatographic or extraction steps for selective titanium isolation.
Given the chemical properties of Ti(IV), effective separation could be based on the sorption of its chloride or oxalate complexes using anion-exchange chromatography or specific sorbents (e.g., ZR resin), as applied to Sc-targets~\cite{Radchenko2016, Benabdallah2023}.
However, applying these methods to calcium-based targets remains largely theoretical and requires specific experimental confirmation.

After purification, the isolated $^{44}$Ti can be immobilised on a solid support to serve as the long-lived parent in a radionuclide generator system.
Typical configurations involve loading $^{44}$Ti onto an anion-exchange resins (e.g., Dowex/AG 1$\times$8) or chemically selective sorbents such as ZR resin, where it remains strongly retained~\cite{Filosofov2010, Radchenko2016, Benabdallah2023, Feng2025}.
The daughter nuclide $^{44}$Sc, generated by electron capture decay~\cite{NuDat3}, can be periodically eluted using weakly acidic solutions (e.g. 0.05–0.1~M HCl), enabling repeated separation cycles with minimal titanium breakthrough~\cite{Filosofov2010, Schmidt2023}.
This approach provides a long-term and cyclotron-independent source of $^{44}$Sc with high apparent molar activity and radionuclidic purity, suitable for radiopharmaceutical applications.

\section{Cyclotrons well-suited for 
\texorpdfstring{$^{45}$Sc(p,2n)$^{44}$Ti}{45Sc(p,2n)44Ti} reaction}

Analysis of the various production pathways demonstrates that the $^{45}$Sc(p, 2n)$^{44}$Ti reaction on natural scandium represents the optimal strategy for establishing the $^{44}$Ti/$^{44}$Sc generator system (see Sect.~\ref{sec:reactions}).
The requirements are a high-flux proton beam above $\sim$15 MeV (ideally 20--30 MeV) and extended irradiation times.
Modern PET/SPECT cyclotrons capable of delivering tens of MeV protons at beam currents ranging from tens to hundreds of $\mu$A are particularly well-suited.

The  Heavy Ion Laboratory of the University of Warsaw (HIL-UW), Poland, operates two cyclotrons used for the production of medical radionuclides, particularly scandium isotopes.
The first is the U-200P cyclotron, designed for the acceleration of heavy ions including $\alpha$ particles, which can reach energies of approximately 32~MeV~\cite{Szkliniarz2015}.
Although this is below the energy required for efficient $^{44}$Ti production, it has been successfully used for the direct production of $^{43}$Sc and $^{44g,m}$Sc~\cite{Szkliniarz2015, Szkliniarz2016, Bilewicz2015, Walczak2015}.
The second is a GE PETtrace cyclotron, delivering 16.5~MeV protons and 8.4~MeV deuterons~\cite{Sitarz2018}.
While the deuteron energy is below the threshold for the $^{45}$Sc(d,3n)$^{44}$Ti reaction ($\sim$15.3~MeV~\cite{Hassan2018}), the available proton beam energy enables partial $^{44}$Ti production near the threshold ($\sim$12.65~MeV~\cite{Daraban2009}) of the $^{45}$Sc(p,2n)$^{44}$Ti channel.
Both beams are currently used for the routine production of $^{43}$Sc and $^{44}$Sc~\cite{Sitarz2016, Sitarz2018, Stolarz2018, Wojdowska2019}.

Preparations are currently ongoing at the HIL to enable $^{44}$Ti production for a $^{44}$Ti/$^{44}$Sc generator designed to calibrate the J-PET scanner.
This technical work includes installing a suitable target and optimising irradiation parameters.
Such a generator would provide a reliable, long-term source of $^{44}$Sc, which is considered an ideal radioisotope for the three-photon imaging technique~\cite{Thirolf2015}, which exploits a high-energy gamma transition to improve image resolution, typically limited by positron range in tissue (see Sect.~\ref{sec:intro}).
Under the available energy window of 16.5$\rightarrow$12~MeV, a 60-hour irradiation with a 10~$\mu$A proton beam would theoretically yield up to 0.299~MBq of $^{44}$Ti, increasing to 0.598~MBq at 20~$\mu$A.
The required target areal density for this configuration is estimated at around 191~mg/cm$^{2}$ (see Table~\ref{tab:thickness}).

iThemba LABS (South Africa) offers a promising infrastructure for the $^{44}$Ti production, with advanced accelerator systems routinely used for large-scale medical radionuclide production.
The K=200 Separated Sector Cyclotron (SSC) accelerates protons, deuterons, $\alpha$ particles, and heavy ions~\cite{Conradie2007}.
Although its maximum proton energy reaches 200~MeV, beams of 66~MeV are typically used for routine radionuclide production.
High-intensity beams of up to 250~$\mu$A can be delivered to the Vertical Beam Target Station (VBTS)~\cite{Conradie2019}.
For $^{44}$Ti production, the proton energy can be reduced to the optimal 20--30~MeV range using energy degraders or stacked targets~\cite{Fantoni1994, Szelecsenyi2005} with beam currents of several hundred $\mu$A.
In parallel, the more recent South African Isotope Facility (SAIF) features a dedicated 70~MeV H$^-$-cyclotron designed exclusively for radioisotope production~\cite{Conradie2019}.
It is capable of generating two independent proton beams simultaneously, each with a current of up to 375~$\mu$A.
Currently, two target stations are operational, with the infrastructure supporting up to four.
Both installations are well-suited for efficient realisation of the $^{45}$Sc(p,2n)$^{44}$Ti reaction.

The RFT-30 cyclotron at the Korea Atomic Energy Research Institute (KAERI), provides a proton beam with energy in the 15--30~MeV range and operates at high beam currents up to 300$~\mu$A~\cite{KAERI}.
Its suitability for $^{44}$Ti production was demonstrated in test experiment using natural scandium targets~\cite{Lee2017}.
A cooled vacuum chamber and inclined target geometry were used to reduce beam density and enhance irradiation efficiency.
The formation of $^{44}$Ti was verified via $\gamma$-spectroscopy, confirming the technical feasibility of using the RFT-30 to establish a $^{44}$Ti/$^{44}$Sc generator based on the $^{45}$Sc(p,2n)$^{44}$Ti reaction.

The ARRONAX (Accelerator for Research in Radiochemistry and Oncology at Nantes Atlantique) facility in Nantes, France, is a high-tech cyclotron specifically designed for the production of medical radionuclides.
It accelerates protons in the 30--70~MeV energy range at beam currents of up to 2$\times$375~$\mu$A~\cite{Haddad2011, Poirier2012}.
ARRONAX supports the use of proton, deuteron, and $\alpha$-particle beams. Energy degraders positioned on the vacuum beam lines allow the proton energy to be reduced to the 10--30~MeV range~\cite{Poirier2012}.
It is actively used for the production of medical radionuclides, including $^{44}$Sc and $^{47}$Sc, and its infrastructure is suitable for $^{44}$Ti synthesis for generator systems.

Table~\ref{tab:facilities} provides a comparative overview of the proton beam energies and currents available at selected cyclotron facilities suitable for $^{44}$Ti production.
It should be noted that many other existing cyclotrons may also be appropriate, provided they deliver proton beams within the optimal energy range and sufficient beam current.

\begin{table*}[!htbp]
    \caption{Proton beam characteristics of selected operational cyclotron facilities suitable for \texorpdfstring{$^{44}$Ti}{44Ti} production via the \texorpdfstring{$^{45}$Sc(p,2n)$^{44}$Ti}{45Sc(p,2n)44Ti} reaction. Listed are nominal proton energies and extracted beam current.}
    \label{tab:facilities}
    \centering
    \begin{tabular}{l|c|c}
    \hline\hline
    Facility & Proton Energy [MeV] & Current [$\mu$A] \\
    \hline
    HIL-UW PETtrace~\cite{Sitarz2018} (Poland) & 16.5 & up to 75 \\
    iThemba LABS K=200 SSC~\cite{Conradie2019} (South Africa) & up to 200 (20--30 on target) & up to 250 \\
    iThemba LABS SAIF~\cite{Conradie2019} (South Africa) & up to 70 (20--30 on target) & 2$\times$375 \\
    KAERI RFT-30~\cite{KAERI} (Korea) & 15--30 & up to 300 \\
    ARRONAX C70~\cite{Haddad2011, Poirier2012} (France) & 30--70 & 2$\times$375 \\
    \end{tabular}
\end{table*}

\section*{Conclusions}

The demand for modern PET and positron imaging technologies requires the development of a stable supply chain for the suitable radionuclides.
In this context, the $^{44}$Ti/$^{44}$Sc generator represents the most robust and practical solution for achieving a decentralised, long-term $^{44}$Sc availability, offering a sustainable platform for routine clinical and research applications.

The $^{45}$Sc(p,2n)$^{44}$Ti reaction was identified as the most favourable production route due to its relatively high cross-section~\cite{Daraban2009, Ejnisman1996, Ditroi2024, Levkovskij1991} (Fig.~\ref{fig:xs_45Sc_p2n}), the monoisotopic nature of natural scandium~\cite{NISTAtomicWeights}, and compatibility with modern cyclotron facilities~\cite{Sitarz2018, Conradie2019, KAERI, Haddad2011} (Table~\ref{tab:facilities}).
Theoretical yield estimates confirm that extended irradiations with proton beams in the 16.5--30~MeV energy range can produce measurable activities of $^{44}$Ti even at moderate beam currents (Fig.~\ref{fig:activity_eob}).
Subsequent chemical processing using advanced extraction resins could ensure high-purity $^{44}$Ti~\cite{Filosofov2010, Radchenko2016, Radchenko2017, Benabdallah2023}, which is essential for a generator with minimal parent breakthrough~\cite{Filosofov2010, Schmidt2023}.

Several existing facilities, including PETtrace at HIL-UW~\cite{Sitarz2018}, the K=200 at iThemba LABS~\cite{Conradie2019}, RFT-30 at KAERI~\cite{Lee2017, KAERI}, and ARRONAX~\cite{Haddad2011, Poirier2012}, have been identified as technically suitable for efficient $^{44}$Ti production.
Successful implementation requires careful optimisation of target cooling and areal density to support high beam currents over extended irradiation periods and ensure high production yields.

The $^{44}$Ti/$^{44}$Sc generator concept is especially attractive when combined with innovative imaging systems such as the plastic scintillator–based J-PET scanner~\cite{Moskal2020, Moskal2014, Moskal2021c, Sharma2023, Das2024, Moskal2025b}, which uses prompt $\gamma$-emission and three-photon coincidence detection to improve spatial resolution~\cite{Das2025}.
Such integration offers the potential to significantly lower the cost of PET imaging, increase reliability, and enable decentralised diagnostic services, particularly in regions with limited medical infrastructure~\cite{Moskal2024a}.
This approach contributes to long-term sustainability and broader clinical adoption of $^{44}$Sc-based diagnostic protocols.

\begin{acknowledgments}
We acknowledge support from the National Science Centre of Poland through grants no. 2021/42/A/ST2/00423, 2021/43/B/ST2/02150, 2022/47/I/NZ7/03112, the SciMat and qLife Priority Research Areas budget under the program Excellence Initiative -- Research University at Jagiellonian University, and the European Union within the Horizon Europe Framework Programme (ERC Advanced Grant POSITRONIUM no. 101199807).
\end{acknowledgments}

\appendix
\section{Evaluation of nuclear reaction cross sections for \texorpdfstring{$^{44}$Ti}{44Ti} production}
\label{app:model_parameters}

Optimising $^{44}$Ti production requires precise knowledge of the reaction cross sections and control over competing nuclear processes.
This appendix compares experimental data from the EXFOR database~\cite{EXFOR} with theoretical evaluations from the TENDL-2023 library~\cite{Koning2019, TENDL2023} and describes attempts to improve their agreement by varying parameters of the TALYS nuclear model~\cite{Koning2023}.

\begin{figure}
    \centering
    \includegraphics[width=\linewidth]{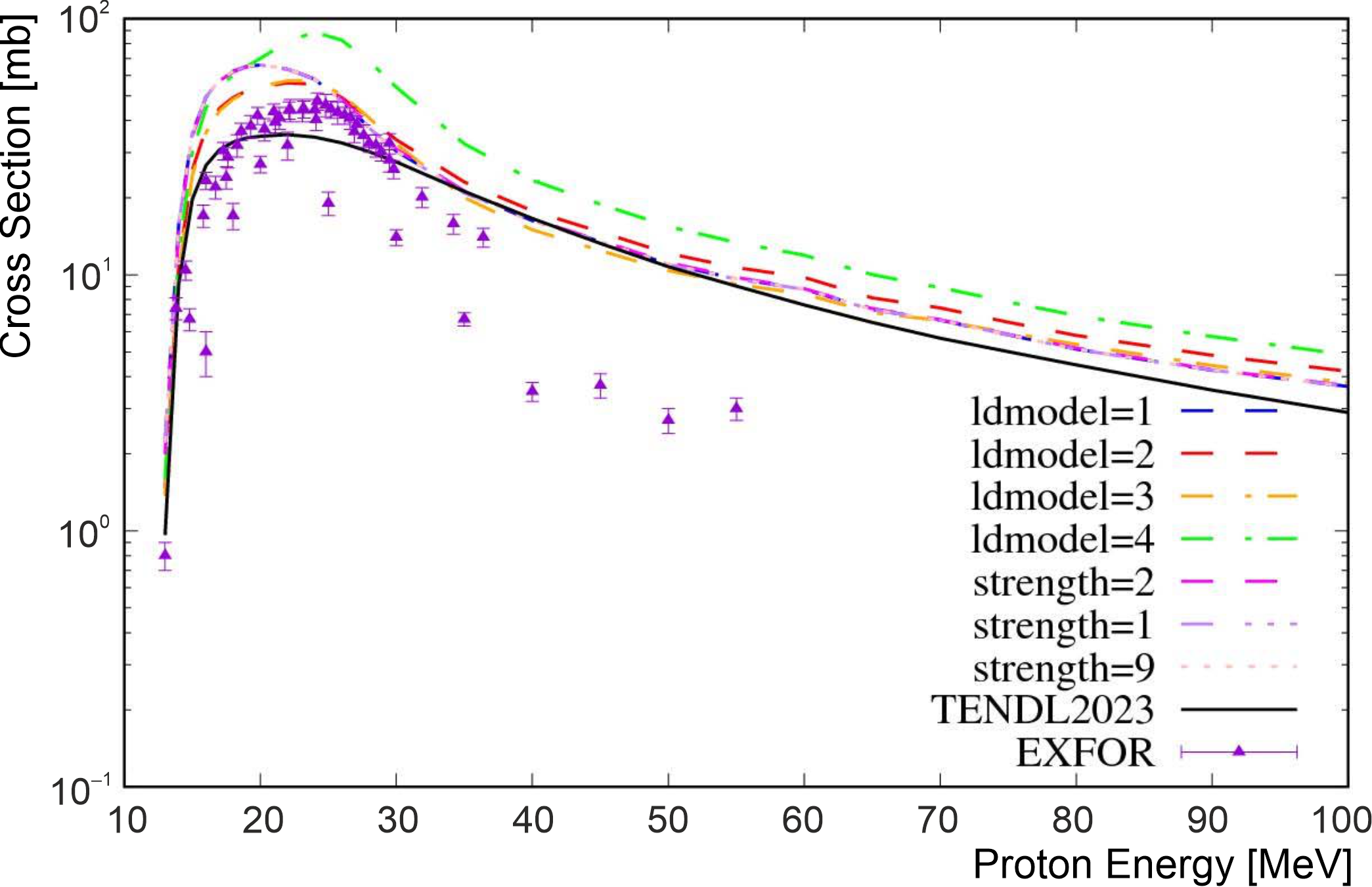}
    \caption{Cross-section for $^{44}$Ti production from proton irradiation of $^{45}$Sc as a function of incident proton energy.
    Experimental data from EXFOR~\cite{Levkovskij1991, Ejnisman1996, Daraban2009, Ditroi2024} are shown as purple triangles.
    The black curve represents evaluated data from TENDL-2023~\cite{TENDL2023}, while coloured lines show TALYS simulations~\cite{Koning2023} using different combinations of level density models (\texttt{ldmodel}~=~1--4) and $\gamma$-ray strength functions (\texttt{strength}~=~1, 2, 9).}
    \label{fig:XS_IK_p45Sc}
\end{figure}

The $^{45}$Sc(p,2n)$^{44}$Ti reaction remains the most widely studied $^{44}$Ti production route due to its favourable cross-section and accessibility of accelerators.
Experimental data~\cite{Levkovskij1991, Ejnisman1996, Daraban2009, Ditroi2024} indicate a broad cross-section maximum of approximately 40~mb around 25~MeV (Fig.~\ref{fig:XS_IK_p45Sc}).
However, TENDL-2023 evaluation~\cite{TENDL2023} systematically underestimate experimental values in the 20--30~MeV energy range, which is most relevant for thick-target irradiations.

To improve the agreement between theory and experiment, a series of TALYS simulations~\cite{Koning2023} was performed by varying key nuclear model parameters.
Specifically, we explored four alternative level density models: the Constant Temperature plus Fermi Gas model (CTM, \texttt{ldmodel}~=~1, TENDL default), the Back-Shifted Fermi Gas model (BFM, \texttt{ldmodel}~=~2), the Generalised Superfluid model (GSM, \texttt{ldmodel}~=~3), and the Skyrme-Hartree-Fock-Bogolyubov model based on tabulated data (HFBM, \texttt{ldmodel}~=~4).
These models govern the statistical population of excited states and are known to affect pre-equilibrium emission probabilities.
In addition, three $\gamma$-ray strength functions were examined: the Kopecky-Uhl Generalized Lorentzian (\texttt{strength}~=~1, TENDL default), the Brink-Axel Lorentzian (\texttt{strength}~=~2), and the Simplified Modified Lorentzian (SMLO, \texttt{strength}~=~9).
These strength functions influence the competition between particle emission and $\gamma$ de-excitation during the decay of the compound nucleus.

While certain parameter combinations affected the overall shape and magnitude of the calculated excitation functions, none of them succeeded in reproducing the experimentally observed cross-section peak (Fig.~\ref{fig:XS_IK_p45Sc}).
This suggests possible limitations in the reaction modelling or input parameters (e.g. optical model potentials) that are not addressed by simple variation of statistical decay models.

\begin{figure}
    \centering
    \includegraphics[width=\linewidth]{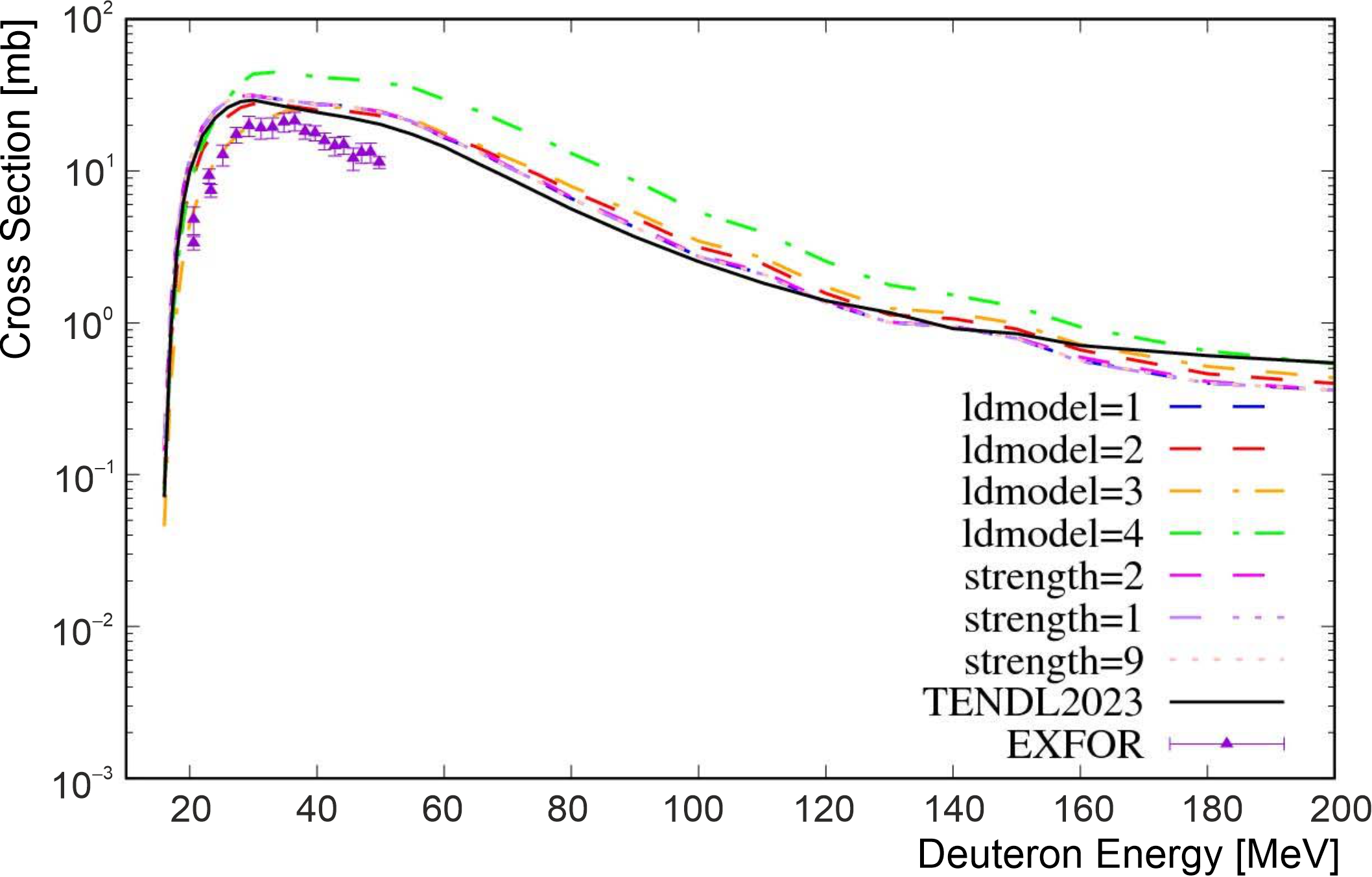}
    \caption{Cross-section for $^{44}$Ti production from deuteron irradiation of $^{45}$Sc as a function of incident deuteron energy.
    Experimental values from EXFOR~\cite{Hermanne2012, Tsoodol2021} are shown as purple triangles.
    The black curve indicates TENDL-2023 evaluations~\cite{TENDL2023}, and coloured lines correspond to TALYS calculations~\cite{Koning2023} using various combinations of level density models (\texttt{ldmodel}~=~1--4) and $\gamma$-ray strength functions (\texttt{strength}~=~1, 2, 9).}
    \label{fig:XS_IK_d45Sc}
\end{figure}

A second potential route involves the $^{45}$Sc(d,3n)$^{44}$Ti reaction.
Experimental cross-section data~\cite{Hermanne2012, Tsoodol2021} and the TENDL-2023 evaluation~\cite{TENDL2023} are shown in Fig.~\ref{fig:XS_IK_d45Sc}.
In contrast to the (p,2n) reaction, the TENDL-2023 evaluation tends to overestimate the cross-section by up to 30\% in the 25--40~MeV energy range, where the maximum reaches approximately 20~mb.
Attempts to improve the agreement between evaluated and experimental data by varying the nuclear model parameters in TALYS~\cite{Koning2023} were also unsuccessful.
Furthermore, this reaction pathway involves over 250 open channels at higher deuteron energies, significantly complicating both the radionuclidic purity and subsequent chemical processing.

\begin{figure*}
    \centering
    \includegraphics[width=0.49\linewidth]{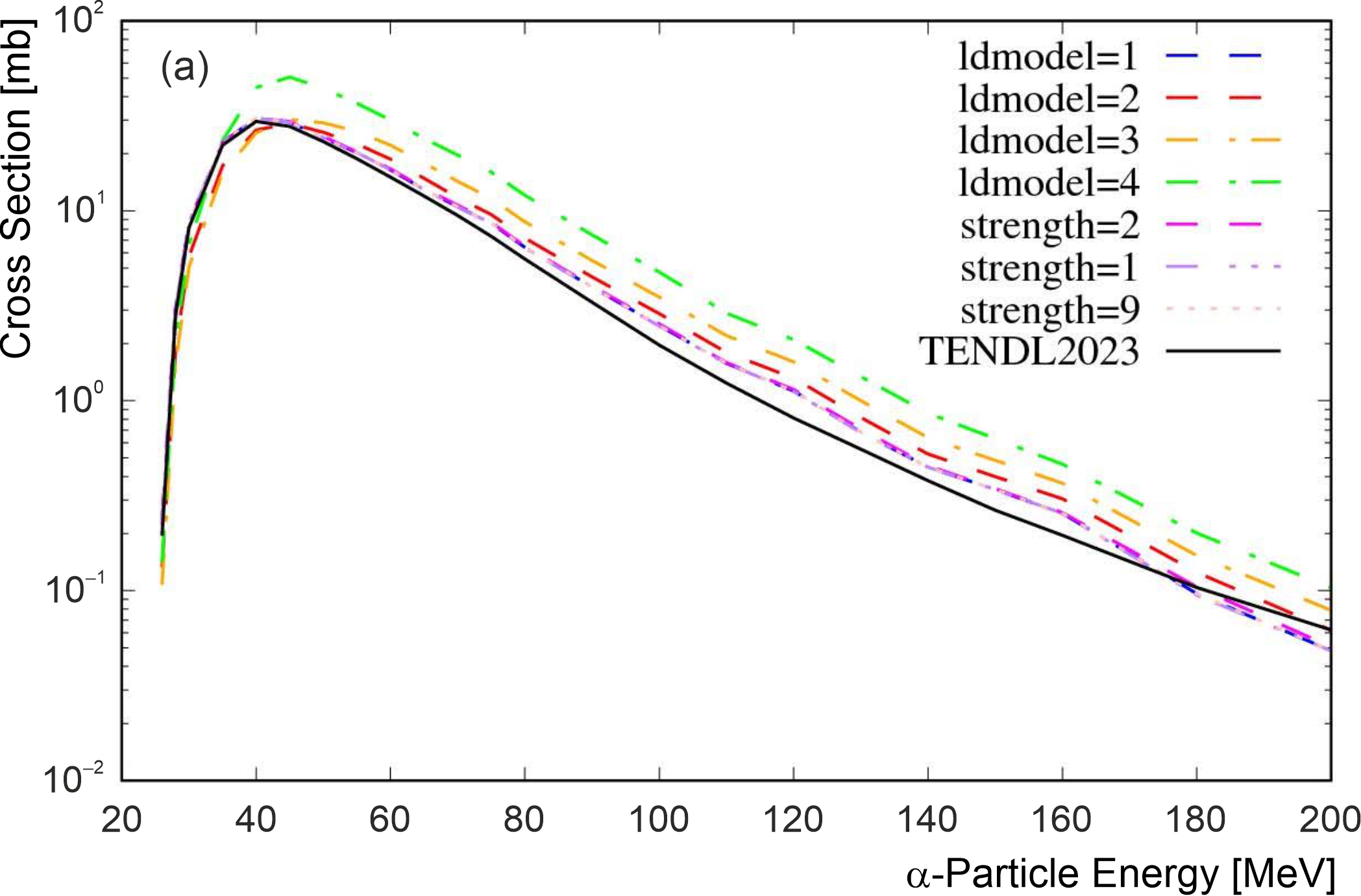}
    \includegraphics[width=0.49\linewidth]{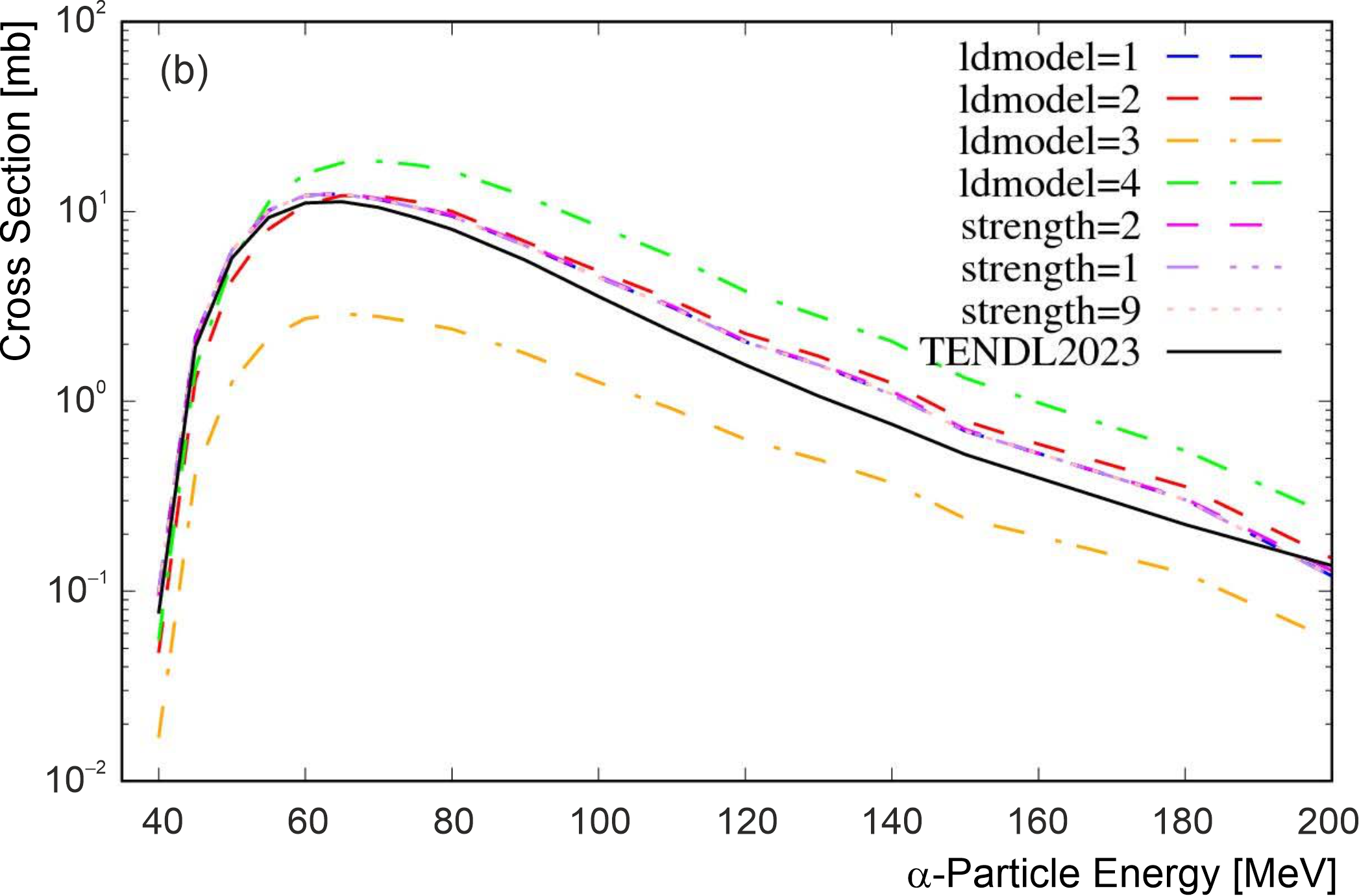}
    \caption{Modelled cross-sections for $^{44}$Ti production via $\alpha$-particle-induced reactions on enriched calcium isotopes:
    (a)~$^{43}$Ca($\alpha$,3n)$^{44}$Ti and (b)~$^{44}$Ca($\alpha$,4n)$^{44}$Ti.
    The black curves show evaluated data from TENDL-2023~\cite{TENDL2023}, while the coloured lines represent TALYS calculations~\cite{Koning2023} using different combinations of level density models (\texttt{ldmodel}~=~1--4) and $\gamma$-ray strength functions (\texttt{strength}~=~1, 2, 9).}
    \label{fig:XS_IK_alpha4344Ca}
\end{figure*}

For $\alpha$-particle-induced reactions on enriched $^{43,44}$Ca isotopes, no experimental cross-section data are currently available.
According to model predictions, the $^{43}$Ca($\alpha$,3n)$^{44}$Ti reaction exhibits a maximum cross-section of approximately 30~mb in the 45--48~MeV range (Fig.~\ref{fig:XS_IK_alpha4344Ca}(a)), while the $^{44}$Ca($\alpha$,4n)$^{44}$Ti reaction reaches 3--12~mb at 60--65~MeV (Fig.~\ref{fig:XS_IK_alpha4344Ca}(b)).
At such energies, however, numerous competing nuclear channels are expected to open, potentially reducing the radionuclidic purity of the resulting $^{44}$Ti.

To evaluate the potential of $^{44}$Ti production via $^7$Li irradiation of $^{40}$Ca, reaction cross-sections were calculated using the EMPIRE nuclear model code~\cite{Herman2007}.
Given the complex reaction mechanism involving both cluster transfer and compound nucleus processes, the calculation considered the cumulative cross-section without resolving individual reaction channels.

The influence of nuclear model parameters on the resulting excitation function was investigated by varying nuclear model parameters implemented in EMPIRE.
We tested five level density prescriptions: the Enhanced Generalised Superfluid Model (EGSM, \texttt{LEVDEN}~=~0), the Generalised Superfluid Model (GSM, \texttt{LEVDEN}~=~1), the Gilbert–Cameron model (GC, \texttt{LEVDEN}~=~2 and 4), and the microscopic Hartree–Fock–Bogoliubov model (HFB, \texttt{LEVDEN}~=~3).
In parallel, we evaluated six $\gamma$-ray strength function models: the Uhl-Kopecki Enhanced Generalised Lorentzian (EGLO, \texttt{GSTRFN}~=~0 and 4), Modified Lorentzian (MLO, \texttt{GSTRFN}~=~1), Generalized Fermi Liquid model (GFL, \texttt{GSTRFN}~=~5), Standard Lorentzian (SLO, \texttt{GSTRFN}~=~6), and the Simplified Modified Lorentzian (SMLO, \texttt{GSTRFN}~=~8).

\begin{figure*}[!htbp]
    \centering
    \includegraphics[width=0.49\linewidth]{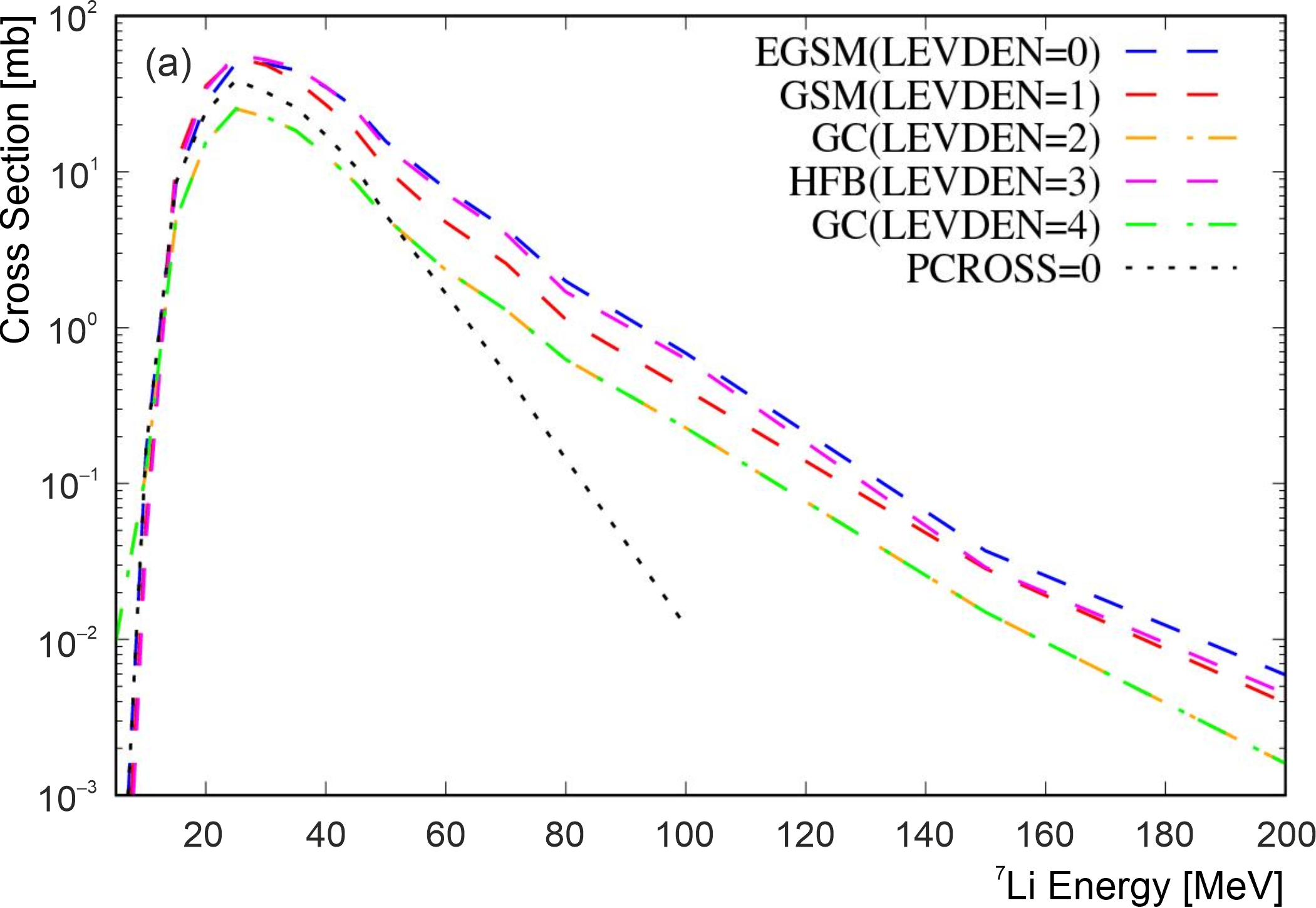}
    \includegraphics[width=0.49\linewidth]{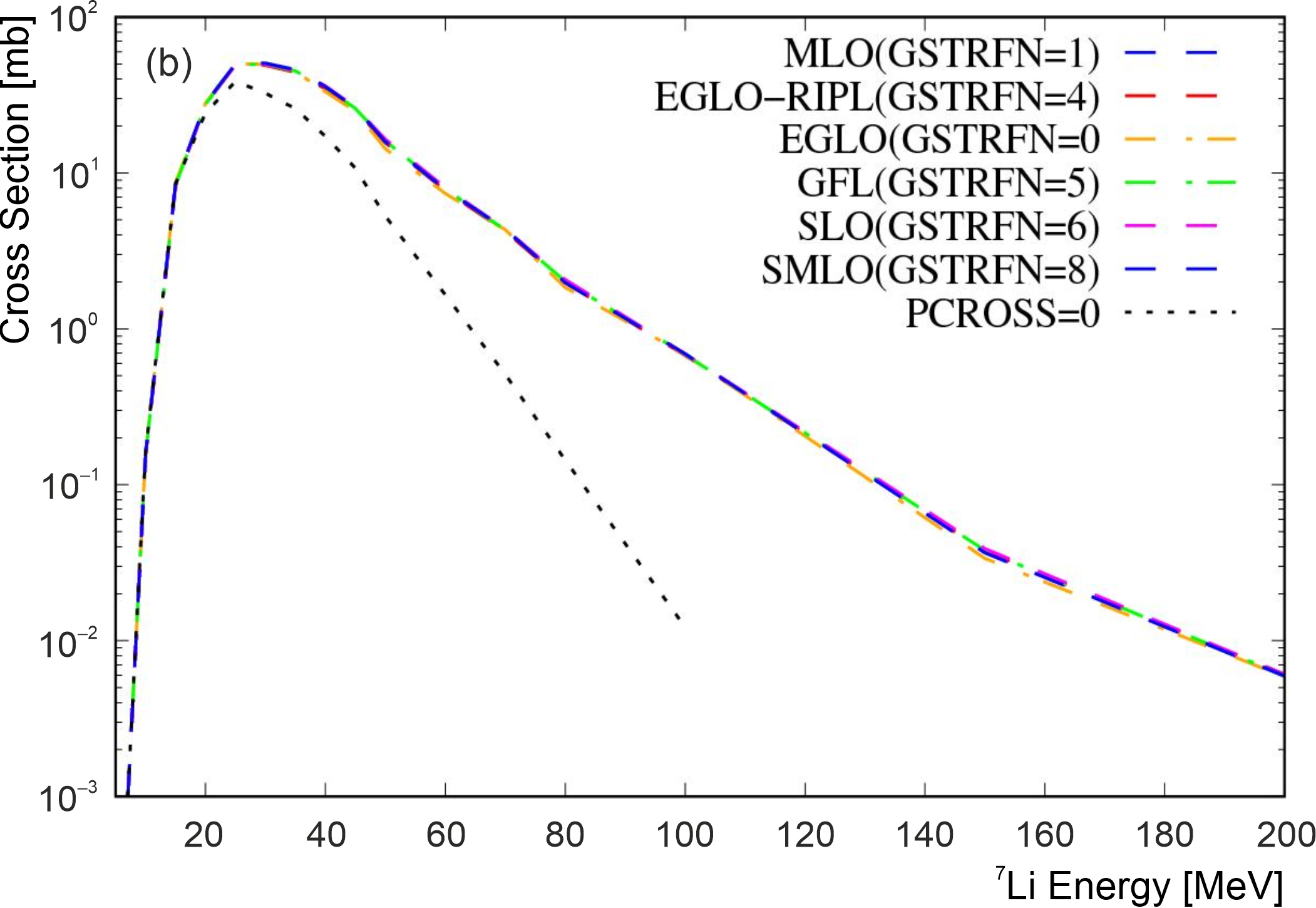}
    \caption{Modelled cumulative cross-sections for $^{44}$Ti production via the $^{40}$Ca($^7$Li,x) reaction.
    (a) Influence of the nuclear level density model (\texttt{LEVDEN}~=~0–4);
    (b) influence of the $\gamma$-ray strength function model (\texttt{GSTRFN}~=~0, 1, 4–6, 8).
    The black dotted line in both panels corresponds to a calculation without pre-equilibrium emission (\texttt{PCROSS}~=~0).}
    \label{fig:XS_IK_7Li_40Ca}
\end{figure*}

The excitation function (Fig.~\ref{fig:XS_IK_7Li_40Ca}) shows a broad maximum of around 45-50~mb at 30--35~MeV.
The choice of level density model has a noticeable impact on both the peak height and high-energy tail, while variations in $\gamma$-ray strength functions yield only minor differences.

\bibliographystyle{apsrev4-2}
\bibliography{references}

\end{document}